\magnification \magstep1
\newdimen\papwidth
\newdimen\papheight
\papwidth=16truecm
\papheight=22truecm
\voffset=0.8truecm
\hoffset=0.1truecm
\hsize=\papwidth
\vsize=\papheight
\advance\vsize   by 0.2truecm
\advance\topskip by 0.2truecm
\newdimen\texpscorrection
\texpscorrection=0truecm 
\def\em{\sl}  

\font\huge=cmbx10 scaled\magstep2
\font\xbold=cmbx10 scaled\magstep1

\font\sub=cmbx10

 at 10truept
\font\tencm=cmcsc10 scaled 900

\font\tenmsb=msbm10
\font\sevenmsb=msbm7
\font\fivemsb=msbm5
\newfam\msbfam
 \textfont\msbfam=\tenmsb \scriptfont\msbfam=\sevenmsb
 \scriptscriptfont\msbfam=\fivemsb
 \def\msb{\fam\msbfam}

\font\teneufm=eufm10
\font\seveneufm=eufm7
\font\fiveeufm=eufm5
\newfam\eufmfam
 \textfont\eufmfam=\teneufm \scriptfont\eufmfam=\seveneufm
 \scriptscriptfont\eufmfam=\fiveeufm

\font\teneurm=eurm10
\font\seveneurm=eurm7
\font\fiveeurm=eurm5
\newfam\eurmfam
 \textfont\eurmfam=\teneurm \scriptfont\eurmfam=\seveneurm
 \scriptscriptfont\eurmfam=\fiveeurm
 \def\eurm{\fam\eurmfam}

\font\teneusm=eusm10
\font\seveneusm=eusm7
\font\fiveeusm=eusm5
\newfam\eusmfam
 \textfont\eusmfam=\teneusm \scriptfont\eusmfam=\seveneusm
 \scriptscriptfont\eusmfam=\fiveeusm
 \def\eusm{\fam\eusmfam}

\font\eightrm=cmr8
\font\eighti=cmmi8
\font\eightsy=cmsy8
\font\eightbf=cmbx8
\font\eighttt=cmtt8
\font\eightsl=cmsl8
\font\eightit=cmti8
\font\sevenit=cmti7

\font\sixrm=cmr6
\font\sixbf=cmbx6
\font\sixi=cmmi6
\font\sixsy=cmsy6
\newfam\truecmr
\newfam\truecmsy

\font\tentruecmr=cmr10
\font\tentruecmsy=cmsy10
\font\eighttruecmr=cmr8
\font\eighttruecmsy=cmsy8
\font\seventruecmr=cmr7
\font\seventruecmsy=cmsy7
\font\sixtruecmr=cmr6
\font\sixtruecmsy=cmsy6
\font\fivetruecmr=cmr5
\font\fivetruecmsy=cmsy5
\textfont\truecmr=\tentruecmr
\scriptfont\truecmr=\seventruecmr
\scriptscriptfont\truecmr=\fivetruecmr
\textfont\truecmsy=\tentruecmsy
\scriptfont\truecmsy=\seventruecmsy
\scriptscriptfont\truecmr=\fivetruecmr
\scriptscriptfont\truecmsy=\fivetruecmsy
\newskip\ttglue
\def \eightpoint{\def\rm{\fam0\eightrm}
\textfont0=\eightrm \scriptfont0=\sixrm \scriptscriptfont0=\fiverm
\textfont1=\eighti \scriptfont1=\sixi   \scriptscriptfont1=\fivei
\textfont2=\eightsy \scriptfont2=\sixsy   \scriptscriptfont2=\fivesy
\textfont3=\tenex \scriptfont3=\tenex   \scriptscriptfont3=\tenex
\textfont\itfam=\eightit  \def\it{\fam\itfam\eightit}%
\textfont\slfam=\eightsl  \def\sl{\fam\slfam\eightsl}%
\textfont\ttfam=\eighttt  \def\tt{\fam\ttfam\eighttt}%
\textfont\bffam=\eightbf  \scriptfont\bffam=\sixbf
\scriptscriptfont\bffam=\fivebf  \def\bf{\fam\bffam\eightbf}%
\tt \ttglue=.5em plus.25em minus.15em
\setbox\strutbox=\hbox{\vrule height7pt depth2pt width0pt}%
\normalbaselineskip=9pt
\let\sc=\sixrm  \let\big=\eightbig  \normalbaselines\rm
\textfont\truecmr=\eighttruecmr
\scriptfont\truecmr=\sixtruecmr
\scriptscriptfont\truecmr=\fivetruecmr
\textfont\truecmsy=\eighttruecmsy
\scriptfont\truecmsy=\sixtruecmsy
}
\catcode`@=12
\fontdimen16\tensy=2.7pt
\fontdimen13\tensy=4.3pt
\fontdimen17\tensy=2.7pt
\fontdimen14\tensy=4.3pt
\fontdimen18\tensy=4.3pt
\fontdimen16\eightsy=2.7pt
\fontdimen13\eightsy=4.3pt
\fontdimen17\eightsy=2.7pt
\fontdimen14\eightsy=4.3pt
\catcode`@=11
\def\footnote#1{\let\@sf\empty 
  \ifhmode\edef\@sf{\spacefactor\the\spacefactor}\/\fi
  #1\@sf\vfootnote{#1}}
\def\vfootnote#1{\insert\footins\bgroup\eightpoint
  \interlinepenalty\interfootnotelinepenalty
  \splittopskip\ht\strutbox 
  \splitmaxdepth\dp\strutbox \floatingpenalty\@MM
  \leftskip\z@skip \rightskip\z@skip \spaceskip\z@skip
  \xspaceskip\z@skip
  \textindent{#1}\footstrut\futurelet\next\fo@t}
\def\fo@t{\ifcat\bgroup\noexpand\next \let\next\f@@t
  \else\let\next\f@t\fi \next}
\def\f@@t{\bgroup\aftergroup\@foot\let\next}
\def\f@t#1{#1\@foot}
\def\@foot{\strut\egroup}
\def\footstrut{\vbox to\splittopskip{}}
\skip\footins=\bigskipamount 
\count\footins=1000 
\dimen\footins=8in 
\catcode`@=12 
\newcount\FIGUREcount \FIGUREcount=0
\newdimen\figcenter
\def\figure #1 #2 #3 #4\cr{\null\ifundefined{fig#1}\global
\advance\FIGUREcount by 1\newdef
fig,#1,{Fig.~\number\count5.\number\FIGUREcount}\fi
\write16{ FIG \number\count5.\number\FIGUREcount: #1}
{\goodbreak\figcenter=\hsize\relax
\advance\figcenter by -#3truecm
\divide\figcenter by 2
\midinsert\vskip #2truecm\noindent\hskip\figcenter
\includegraphics{#1}\vskip 0.8truecm\noindent
\vbox{\eightpoint\noindent
{\bf\fig(#1)}: #4}\endinsert}}
\def\figurewithtex #1 #2 #3 #4 #5\cr{\null\ifundefined{fig#1}\global
\advance\FIGUREcount by 1\newdef
fig,#1,{Fig.~\number\count5.\number\FIGUREcount}\fi
\write16{ FIG \number\count5.\number\FIGUREcount: #1}
{\goodbreak\figcenter=\hsize\relax
\advance\figcenter by -#4truecm
\divide\figcenter by 2
\midinsert\vskip #3truecm\noindent\hskip\figcenter
\includegraphics{#1}{\hskip\texpscorrection\input #2 }
\vskip 0.8truecm\noindent \vbox{\eightpoint\noindent
{\bf\fig(#1)}: #5}\endinsert}}
\def\fig(#1){\ifundefined{fig#1}\global
\advance\FIGUREcount by 1\newdef
fig,#1,{Fig.~\number\count5.\number\FIGUREcount}
\fi
\csname fig#1\endcsname\relax}
\def\period{\unskip.\spacefactor3000 { }}
\newbox\nobox
\newbox\bybox
\newbox\paperbox
\newbox\yrbox
\newbox\jourbox
\newbox\pagesbox
\newbox\volbox
\newbox\preprintbox
\newbox\toappearbox
\newbox\bookbox
\newbox\bybookbox
\newbox\publisherbox
\newbox\inprintbox
\def\refclear{
   \setbox\nobox=\null        \gdef\isno{F}
   \setbox\bybox=\null        \gdef\isby{F}
   \setbox\paperbox=\null     \gdef\ispaper{F}
   \setbox\yrbox=\null        \gdef\isyr{F}
   \setbox\jourbox=\null      \gdef\isjour{F}
   \setbox\pagesbox=\null     \gdef\ispages{F}
   \setbox\volbox=\null       \gdef\isvol{F}
   \setbox\preprintbox=\null  \gdef\ispreprint{F}
   \setbox\toappearbox=\null  \gdef\istoappear{F}
   \setbox\inprintbox=\null   \gdef\isinprint{F}
   \setbox\bookbox=\null      \gdef\isbook{F}  \gdef\isinbook{F}
   \setbox\bybookbox=\null    \gdef\isbybook{F}
   \setbox\publisherbox=\null \gdef\ispublisher{F}
}
\def\ref{\refclear\bgroup}
\def\no   {\egroup\gdef\isno{T}\setbox\nobox=\hbox\bgroup}
\def\by   {\egroup\gdef\isby{T}\setbox\bybox=\hbox\bgroup}
\def\paper{\egroup\gdef\ispaper{T}\setbox\paperbox=\hbox\bgroup}
\def\yr{\egroup\gdef\isyr{T}\setbox\yrbox=\hbox\bgroup}
\def\jour{\egroup\gdef\isjour{T}\setbox\jourbox=\hbox\bgroup}
\def\pages{\egroup\gdef\ispages{T}\setbox\pagesbox=\hbox\bgroup}
\def\vol{\egroup\gdef\isvol{T}\setbox\volbox=\hbox\bgroup\bf}
\def\preprint{\egroup\gdef
\ispreprint{T}\setbox\preprintbox=\hbox\bgroup}
\def\toappear{\egroup\gdef
\istoappear{T}\setbox\toappearbox=\hbox\bgroup}
\def\inprint{\egroup\gdef
\isinprint{T}\setbox\inprintbox=\hbox\bgroup}
\def\book{\egroup\gdef\isbook{T}\setbox\bookbox=\hbox\bgroup\em}
\def\publisher{\egroup\gdef
\ispublisher{T}\setbox\publisherbox=\hbox\bgroup}
\def\inbook{\egroup\gdef\isinbook{T}\setbox\bookbox=\hbox\bgroup\em}
\def\bybook{\egroup\gdef\isbybook{T}\setbox\bybookbox=\hbox\bgroup}
\def\endref{\egroup \sfcode`.=1000
 \if T\isno  \item{[\unhbox\nobox\unskip]}
     \else  \noindent    \fi
 \if T\isby    \unhbox\bybox\unskip: \fi
 \if T\ispaper \unhbox\paperbox\unskip\period \fi
 \if T\isbook {\it\unhbox\bookbox\unskip}\if T\ispublisher, \else.
\fi\fi
 \if T\isinbook In {\it\unhbox\bookbox\unskip}\if T\isbybook,
\else\period \fi\fi
 \if T\isbybook  (\unhbox\bybookbox\unskip)\period \fi
 \if T\ispublisher \unhbox\publisherbox\unskip \if T\isjour, \else\if
T\isyr \  \else\period \fi\fi\fi
 \if T\ispreprint Pre\-print\period \fi
 \if T\isjour    \unhbox\jourbox\unskip\ \fi
 \if T\istoappear (to appear)\period \fi
 \if T\isvol     \unhbox\volbox\unskip\if T\ispages, \else\ \fi\fi
 \if T\ispages   \unhbox\pagesbox\unskip\  \fi
 \if T\isyr      (\unhbox\yrbox\unskip)\period \fi
 \if T\isinprint (in print)\period \fi
\filbreak
}
\def\\{\backslash}

\def\onehalf{{\textstyle{1\over2}}}
\def\onefourth{{\textstyle{1\over4}}}
\def\oneeighth{{\textstyle{1\over8}}}

\def\sign{\mathop{\rm sign}}
\def\...{{$\cdot\cdot\cdot$}}
\footline{\ifnum\pageno=0\hss\else\hss\tenrm\folio\hss\fi}

\def\mean{{\rm I\kern-.18em E}}

\def\integer{{\msb Z}}
\def\real{{\msb R}}
\def\complex{{\msb C}}
\def\identity{{\rm 1\kern-.25em l}}
\def\torus{{\msb T}}
\def\dn{|\kern-.08em\|}

\def\AA{{\cal A}}
\def\BB{{\cal B}}
\def\CC{{\cal C}}
\def\DD{{\cal D}}

\def\FF{{\cal F}}
\def\GG{{\cal G}}

\def\II{{\cal I}}
\def\JJ{{\cal J}}

\def\LL{{\cal L}}

\def\OO{{\cal O}}
\def\PP{{\cal P}}
\def\QQ{{\cal Q}}
\def\RR{{\cal R}}

\def\TT{{\cal T}}

\def\ZZ{{\cal Z}}
\def\sK{{\eusm K}}
\def\sB{{\eusm B}}
\def\sO{{\eusm O}}
\def\rp{{\eurm p}}
\def\rq{{\eurm q}}

\def\Im{\mathop{\rm Im}\nolimits}
\def\sump{\mathop{\sum{\vphantom{\sum}}^\prime}}
\def\imun{{\rm i}}
\count5=0       
\count6=1       
\count7=1       
\count8=1       
\def\title#1\par{\centerline{\huge #1}}
\def\abstract#1\par{\noindent{\sub Abstract.} #1 \par}
\def\proof{\medskip\noindent{\bf Proof.\ }}

\def\qed{\hfill\vskip 1pt
         \line{\hfill\vrule height 1.8ex width 2ex depth +.2ex
               \ \ \ \ \ \ }\bigskip}

\def\notation{\medskip\noindent{\bf Notations.\ }}

%
\def\equs(#1){(\csname e#1\endcsname)}
\def\ifundefined#1{\expandafter\ifx\csname#1\endcsname\relax}
%
\def\equation(#1){\ifx\draft\undefined \eqno(\tag(#1))
                  \else \eqno{{\rm #1}(\tag(#1))}\fi}
\def\equ(#1){\ifundefined{e#1}$\spadesuit$#1\else
(\csname e#1\endcsname)\fi}
\def\equat(#1){\ifx\draft\undefined \hfil(\tag(#1))
               \else \hfil{{\rm #1}(\tag(#1))}\fi}
\def\aequation(#1){\ifx\draft\undefined \eqno(\atag(#1))
                   \else \eqno{{\rm #1}(\atag(#1))}\fi}
\def\aequat(#1){\ifx\draft\undefined \hfil(\atag(#1))
                \else \hfil{{\rm #1}(\atag(#1))}\fi}
\def\equationp(#1){\ifx\draft\undefined \eqno(\csname e#1\endcsname')
                   \else \eqno{{\rm #1}(\csname e#1\endcsname')}\fi}
\def\equp(#1){\ifundefined{e#1}$\spadesuit$#1\else
(\csname e#1\endcsname')\fi}

\def\newdef #1,#2,#3 {\ifundefined{#1#2}
    \expandafter\xdef\csname #1#2\endcsname{#3}\else
    \write16{!!!!!doubly defined #1,#2}\fi}
\def\tag(#1){\number\count5.\number\count6
    \newdef e,#1,\number\count5.\number\count6
    \global\advance\count6 by 1}
\def\atag(#1){{\rm A.}\number\count6
    \newdef e,#1,A.\number\count6
    \global\advance\count6 by 1}
\def\claim #1(#2) #3\par{
    \vskip.1in\medbreak\noindent
    {\ifx\draft\undefined\else\llap{(#2)}\fi
    \bf #1\ \number\count5.\number\count7.\ }{\sl #3}\par
    \newdef c,#2,#1~{\number\count5.\number\count7}
    \global\advance\count7 by 1
    \ifdim\lastskip<\medskipamount
    \removelastskip\penalty55\medskip\fi}
\def\claimrm #1(#2) #3\par{
    \vskip.1in\medbreak\noindent
    {\ifx\draft\undefined\else\llap{(#2)}\fi
    \bf #1\ \number\count5.\number\count7.\ }{\rm #3}\par
    \newdef c,#2,#1~{\number\count5.\number\count7}
    \global\advance\count7 by 1
    \ifdim\lastskip<\medskipamount
    \removelastskip\penalty55\medskip\fi}
\def\clm(#1){\ifundefined{c#1}$\spadesuit$#1\else
             \csname c#1\endcsname\fi}
\def\section#1\par{\vskip0pt plus.3\vsize\penalty-75
    \vskip0pt plus -.3\vsize\bigskip\bigskip
    \global\advance\count5 by 1
    \message{#1}\leftline
     {\xbold \number\count5.\ #1}
    \count6=1
    \count7=1
    \count8=1
    \nobreak\smallskip\noindent}
\def\sectionnonr#1\par{\vskip0pt plus.3\vsize\penalty-75
    \vskip0pt plus -.3\vsize\bigskip\bigskip
    \global\advance\count5 by 1
    \message{#1}\leftline
     {\xbold #1}
    \count6=1
    \count7=1
    \count8=1
    \nobreak\smallskip\noindent}
\def\sec(#1){\ifundefined{s#1}$\spadesuit$#1
             \else Section \csname s#1\endcsname\fi}
\def\subsection#1\par{\vskip0pt plus.2\vsize\penalty-75
    \vskip0pt plus -.2\vsize\bigskip\bigskip
    \message{#1}\leftline{\sub
    \number\count5.\number\count8.\ #1}
    \global\advance\count8 by 1
    \nobreak\smallskip\noindent}
\footline{
 \ifx\draftfoot\undefined {\hss \tencm\folio\hss}
 \else
   \hbox to 0pt{\sevenit Draft version --
                \the\day.\the\month.\the\year\hss}
   {\hss \tencm\folio\hss}
 \fi}

\newdimen\normalskip
\newdimen\normalindent
\newdimen\greatbaselineskip
\normalskip=5pt
\normalindent=20pt
\greatbaselineskip=5.25mm
\normalbaselineskip=\baselineskip

\parskip=\normalskip
\pageno=1
\hsize=16.0truecm\vsize=22.5truecm
\null\vskip 1cm

\title Renormalization Group and the Melnikov

\title Problem for PDE's

\vskip 1cm
\centerline
{\bf
Jean Bricmont$^{1,}$\footnote
{$^\dagger$}{Partially supported by ESF/PRODYN.},
Antti Kupiainen$^{2,}$\footnote
{$^\ddagger$}{Partially supported by EC grant FMRX-CT98-0175.},
Alain Schenkel$^{2}$
}
\vskip 0.5cm
{\noindent\eightrm\baselineskip=10pt
\raise 1mm\hbox{$\scriptstyle 1$}$\,$UCL, FYMA, 2 chemin du Cyclotron,
B-1348 Louvain-la-Neuve, Belgium\hfill\break
\raise 1mm\hbox{$\scriptstyle 2$}$\,$Department of Mathematics,
Helsinki University, P.O.~Box~4, 00014 Helsinki, Finland\hfill\break
\par}
\vskip 1.2cm
\abstract
We give a new proof of persistence of quasi-periodic, low dimensional
elliptic tori in infinite dimensional systems. The proof is based on a
renormalization group iteration that was developed recently in
[BGK] to
address the standard KAM problem, namely, persistence of invariant
tori of maximal dimension in finite dimensional, near
integrable systems.
Our result covers situations in which the so called normal frequencies
are multiple. In particular, it provides a new proof of
the existence of
small-amplitude, quasi-periodic solutions of nonlinear wave equations
with periodic boundary conditions.

\section Introduction

In this paper, we address the persistence problem of quasi-periodic,
low dimensional, elliptic tori in infinite dimensional systems.
A typical example that we will consider is the nonlinear wave equation
(NLW) on a bounded interval,
$$
\partial_t u=\partial_x^2 u-Vu+f(u),
\equation(NLWintro)
$$
with Dirichlet or periodic boundary conditions and $f(u)=\OO(u^3)$.
The first results concerning the existence of quasi-periodic solutions
of \equ(NLWintro) were obtained independently by Kuksin, P\"oschel and
Wayne, [K, P1, W]. They extended
to infinite dimensional Hamiltonian
systems Eliasson's proof, [E], of the
so called Melnikov problem, {i.e.}, the persistence of
elliptic invariant tori of dimension lower than the number of
degrees of freedom.
Based on the Kolmogorov-Arnold-Moser (KAM) approach,
these results were restricted, however, to Dirichlet or Neumann
boundary conditions and to specific classes of
potential $V$ excluding, in particular, the case $V=Const.$
In [P2], P\"oschel covered the case of constant potentials by
exploiting the existence of a Birkhoff normal form
for the Hamiltonian of \equ(NLWintro).
The normal form allowed him to control the torus
frequencies via amplitude-frequency modulation, and therefore to
dispense with outer parameters provided by an adjustable
potential $V(x)$.
This approach was
applied in [KP] to the persistence of quasi-periodic solutions for
the nonlinear Schr\"odinger equation (NLS) subject to Dirichlet
(or Neumann) boundary conditions.

The case of periodic boundary conditions is more delicate due to the
fact that the eigenvalues of the Sturm-Liouville operator
$L=-d^2/dx^2+V$
are degenerate.
This leads to resonances between pairs of
frequencies corresponding to motion in directions normal
to the torus (the so called normal frequencies).
These additional resonances prevents one from controlling
quadratic terms
in the Hamiltonian of the system and do not seem to be addressable by
KAM techniques. (This difficulty also appears in finite-dimensional
Melnikov situations.) Developing new techniques based on the
Lyapunov-Schmidt method, Craig and Wayne proved in [CW] persistence of
periodic solutions of the NLW with
periodic boundary conditions. Later, their approach was significantly
improved by
Bourgain in [B1-2] who constructed quasi-periodic
solutions of the NLW and NLS with periodic boundary conditions.
Most notably, it is shown in [B2] that solutions of
this type can be constructed, in particular,
for the NLS on two-dimensional domains.
The usual Melnikov nonresonance condition reads,
with $\omega\in\real^d$ and $\mu\in\real^n$ denoting the torus and,
respectively, the normal frequencies ($n$ is possibly infinite),
$$
\langle k,\omega\rangle+\langle l,\mu\rangle\not=0,
\quad k\in\integer^d,\,l\in\integer^n
\ {\rm with}\ |k|+|l|\not=0,\,|l|\leq2.
\equation(Melcond)
$$
In Bourgain's approach and
at the price of a considerable technical effort,
condition \equ(Melcond) is reduced
to
$$
\langle k,\omega\rangle+\mu_s\not=0,
\quad k\in\integer^d,\ s=1,\dots,n,
$$
i.e., all
nonresonance conditions on pairs of normal frequencies are absent.
More recently, Chierchia and You, see [Y,CY], showed that persistence
of quasi-periodic solutions of the NLW with periodic boundary
conditions is tractable by KAM techniques. Their nonresonance
condition,
$$
\langle k,\omega\rangle+\langle l,\mu\rangle\not=0,
\quad k\in\integer^d\setminus\{0\},\,l\in\integer^n
\ {\rm with}\ |l|\leq2,
\equation(WMelcond)
$$
is weaker than \equ(Melcond), but stronger than Bourgain's condition.
However,
for reasons related to the availability of a normal
form mentioned above,
they are unable to cover the
case of constant potential $V$.
In the present paper, we give a new proof of Bourgain's result for the
NLW with periodic boundary conditions.
To this end, we will use a renormalization group procedure
recently developed
in [BGK] for standard KAM problems.
The nonresonance condition that we will impose is the same as
Chierchia and You's condition, but our technique could in principle
accommodate Bourgain's conditon.

In order to describe our result further, we start by specifying
the infinite dimensio\-nal Hamiltonians we will consider.
For $d_k$, $k\geq1$, a sequence of
strictly positive integers uniformly bounded by some $\bar d<\infty$,
let $\RR^\infty$ denote
the set of infinite sequences
$x=(x_1,x_2,\dots)$ with
$x_k\in\real^{d_k}$.
For an integer $d\geq1$, let
$\PP=\torus^d\times\real^d\times\RR^\infty\times\RR^\infty$
where $\torus^d$ is the torus $\real^d/(2\pi\integer^d)$.
Deno\-ting
the coordinates in $\PP$ by $(\phi,I,x,y)$ and
endowing $\PP$ with the symplectic structure $d\phi\wedge
dI+dx\wedge dy$,
we consider perturbations of integrable Hamiltonians of the form
$$
H(\phi,I,x,y)=\omega\cdot I+{\onehalf}I\cdot gI
+\onehalf\sum_{k\geq1}\bigl(\mu_k^2|x_{k}|^2+|y_{k}|^2\bigr)
+\lambda U(\phi,I,x),
\equation(hamiltonian)
$$
where $\mu_k\in\real$, $k\geq1$,
$\omega\in\real^d$, and $g$ is a real symmetric, invertible
$d\times d$ matrix.
Above, $|v|^2$ for $v\in\real^m$ denotes $\sum_{i=1}^mv_i^2$.
The Hamiltonian flow generated by \equ(hamiltonian) is
given by the equations of motion
$$
\dot I=-\lambda\partial_\phi U\,,\quad
\dot\phi=\omega+gI+\lambda\partial_I U,
\equation(eqmot1)
$$
and
$$
\ddot x_{k}=-\mu_k^2x_{k}-\lambda\partial_{x_{k}}U\,
\equation(eqmot2)
$$
For $\lambda=0$ and the initial condition
$I^0=\phi^0=x^0=y^0=0$,
the flow
$\phi(t)=\omega t$, $I(t)=0$, and $x(t)=0$,
is quasi-periodic and spans a
$d$-dimensional torus in
$\torus^d\times\real^d\times\RR^\infty\times\RR^\infty$.
In order to study the case for which the perturbation is turned on,
we consider a quasi-periodic solution of the form
$$
(\phi(t),I(t),x(t))=(\omega t+\Phi(\omega t),J(\omega t),
Z(\omega t)).
$$
Then, \equ(eqmot1) and \equ(eqmot2) require that
$\TT\equiv(\Phi,J,Z):\torus^d\rightarrow
\real^d\times\real^d\times\RR^\infty$
satisfies the equation
$$
\DD\TT(\varphi)=-\lambda\partial U(\varphi+\Phi(\varphi),
J(\varphi),Z(\varphi)),
\equation(eqmot3)
$$
where $\partial=(\partial_\phi,\partial_I,\partial_x)$ and, setting
$$
\mu\equiv{\rm diag}(\mu_1\identity_{d_1},\mu_2\identity_{d_2},\dots),
\equation(defmu)
$$
together with $D\equiv\omega\cdot\partial_\phi$,
$$
\DD=\pmatrix{0 & D & 0\cr
            -D & g & 0\cr
             0 & 0 & D^2+\mu^2\cr
}.\equation(operator)
$$
Note that if $\TT$ is a solution of equation \equ(eqmot3), then so is
$\TT_\beta$ for $\beta\in\real^d$, where
$$
\TT_\beta(\varphi)=\TT(\varphi-\beta)-(\beta,0,0).
\equation(transinv)
$$
We now state the two hypothesis under which
we shall prove existence of a solution $\TT$ of equation \equ(eqmot3),
first introducing
the following family of Banach spaces $\RR^\infty_s, s\in\real$,
$$
\RR^\infty_s=\{Z\in\RR^\infty\ |\ |Z|_s
\equiv\sum_{k\geq1}k^s|Z_k|_{\real^{d_k}}<\infty\}.
\equation(defRRs)
$$

\noindent{\bf(H1) Asymptotics of eigenvalues.} The sequence
$\{\mu_k\}_{k\geq1}$ satisfies $\mu_k>0$ and $\mu_k\not=\mu_l$
for all $k\not=l\geq1$, and
there exist $\gamma\geq1$ and $c>0$ such that
$$
{\mu_k}\geq ck^\gamma
\quad{\rm for\ all}\quad k\geq1.
\equation(asymptotics1)
$$
Furthermore, if $\gamma>1$ then
$$
{\mu_{k'}-\mu_{k}}\geq c(k'^\gamma-k^\gamma)
\quad{\rm for\ all}\quad k'>k\geq1.
\equation(asymptotics2)
$$
If $\gamma=1$, then there exist constants $\xi>0$ and $c_l>0$
such that
$$
\mu_{k'}-\mu_{k}=c_l(1+\OO(k^{-\xi}))
\quad{\rm for\ all}\quad k'-k=l\geq1.
\equation(asymptotics3)
$$

\noindent{\bf(H2) Regularity of the perturbation.}
The map
$(\phi,I,x)\mapsto U(\phi,I,x)$
is assumed to be real analytic in $\phi\in\torus^d$ and
real analytic in $I$ and $x$ in a
neighborhood of the origin of $\real^d$ and $\RR_0^\infty$.
In addition, we assume
that there exist an $s>0$ and a $\xi>0$ such that
for some $\sO_I\subset\real^d$ and
$\sO_x\subset\RR^\infty_s$ neighborhoods of the origin,
the gradient $\partial_xU$
is bounded
as a map from $\torus^d\times\sO_I\times\sO_x$
to $\RR^\infty_{s+\xi-\gamma}$.
In the sequel, we will often use the short notation
$s'\equiv s+\xi-\gamma$.

\claim Theorem(mainthm)
Let $\{\mu_k\}$ satisfy (H1) and $U$ satisfy (H2).
Then, there exists a set
$\Omega^*=\Omega^*(U,\mu)\subset\real^d$ such that for
$\omega\in\Omega^*$,
equation \equ(eqmot3) has a unique
solution (up to translations \equ(transinv)) which is
real analytic in $\lambda$ and $\phi$ provided that
$|\lambda|$ is small enough. Furthermore,
for all bounded $\Omega\subset\real^d$
the set $\Omega^*$ of
admissible frequencies satisfies
${\rm meas}(\Omega\setminus\Omega^*)\rightarrow0$ as
$\lambda\rightarrow0$.

The proof of \clm(mainthm) is based on an inductive procedure
developed in [BGK] for standard KAM problems.
This renormalization group iteration can be viewed as an iterative
resummation of the Lindsedt series, as is explained
in more details in
[BGK], and was directly inspired by the quantum field theory analogy
with KAM problems forcefully emphasized by Gallavotti {\it et al.}
[G, GGM].
Melnikov type problems require to deal with the additional resonances
arising from the normal frequencies $\mu_k$, and
the goal of the
present paper is to explain how the procedure of [BGK] can be applied
in such cases.
In contrast to standard KAM problems,
the set $\Omega^*$ of admissible frequencies depends
for Melnikov type problems
on the perturbation $U$.
In our approach, this dependence expresses itself by the fact that
under iteration, the normal
frequencies are renormalized in a $U$-dependent way and that
the set $\Omega^*$ is defined according to the renormalized normal
frequencies.
As usual, the set $\Omega^*$ is constructed in such a way that
nonresonance
conditions are fulfilled in order for the inductive
scheme to converge.
Our scheme is technically simplified if one imposes
nonresonance condition of the form \equ(WMelcond), {i.e.},
conditions involving pairs of normal frequencies.
Hypothesis (H1) ensures that $\Omega^*$ has large measure under these
conditions, and
hypothesis (H2) ensures that the asymptotic properties of the normal
frequencies stated in (H1) are preserved under renormalization.
The requirement $\xi>0$ is needed both in (H1) when $\gamma=1$,
and, for $\gamma>1$, in (H2) in order to cover the case of
degenerate normal frequencies (more precisely the case where $d_k>1$
for infinitely many $k$).
In Section~2, we show how \clm(mainthm) provides a proof of the
existence of quasi-periodic solutions of the 1D NLW
with periodic boundary conditions. In particular, $\gamma=1$ in
(H1) and we will see
that (H2) is satisfied with $\xi=1$.
In contrast, one has for the 1D NLS
$\gamma=2$ and $\xi=0$. Thus, the scheme presented here
only applies to NLS with
Dirichlet boundary conditions (namely $d_k=1$ for all $k$) or to the
persistence of periodic solutions of NLS (namely $d=1$).
In order to cover the other situations, one must
be able to dispense with nonresonance conditions involving certain
pairs of
normal frequencies.

The remainder of the paper is organized as follows.
Section~2 is devoted to the NLW.
In Section~3 we
explain the renormalization group scheme that will be used to prove
\clm(mainthm). Section~4 is devoted to
the definition of the spaces we will consider.
In Section~5, we
state some crucial inductive bounds, which will be
shown to hold in Section~6.
Section~7 is concerned with the
measure estimate of $\Omega^*$, whereas
the proof of \clm(mainthm) is carried out in Section~8. Finally, we
have collected in the appendix some technical and
intermediary results.

\section The 1D Wave Equation

In this section,
we show how \clm(mainthm) implies the existence of
small amplitude quasi-periodic solutions of nonlinear
1D wave equations
of the form
$$
\partial_t^2u=\partial_x^2u-mu-f(u),
\equation(nlw)
$$
$t>0$$, x\in[0,2\pi]$, with periodic boundary conditions
$u(0,t)=u(2\pi,t)$, $\partial_tu(0,t)=\partial_tu(2\pi,t)$.
Here, $m>0$ is a real parameter and $f$ is a real analytic function
of the form $f(u)=u^3+\OO(u^4)$.
For $f\equiv0$, equation \equ(nlw) becomes
$$
\partial_t^2u=\partial_x^2u-mu\equiv-Lu.
\equation(linw)
$$
The operator $L$ with periodic boundary conditions admits
a complete orthonormal basis of eigenfunctions
$\psi_n\in L^2([0,2\pi])$,
$n\in\integer$, with corresponding eigenvalues
$$
\zeta_n=n^2+m,
\equation(eigen1)
$$
if one sets $\psi_0=1/\sqrt{2\pi}$ and for $n\geq1$,
$$
\psi_n(x)={1\over\sqrt{\pi}}\cos(nx),\quad
\psi_{-n}(x)={1\over\sqrt{\pi}}\sin(nx).
\equation(vbasis)
$$
Every solution of the linear wave equation \equ(linw) can be
written as a superposition of the basic modes $\psi_n$, namely, for
$\II$ any subset of $\integer$ and $\mu_n\equiv\sqrt{\zeta_n}$,
$$
u(x,t)=\sum_{n\in\II}a_n\cos(\mu_nt+\theta_n)\psi_n(x),
\equation(supersol)
$$
with amplitudes $a_n>0$ and initial phases $\theta_n$.
Regarding existence of solutions for the nonlinear wave equation
\equ(nlw), we will prove the
\claim Theorem(thmnlw)
Let $1\leq d<\infty$ and $\II=\{n_1,\dots,n_d\}\subset\integer$
satisfying $|n_i|\not=|n_j|$ for
$i\not=j$. Then, for $\lambda>0$ small enough there is
a set $\AA\subset\{a=(a_1,\dots,a_d)\,|\,0<a_i<\lambda\}$ of
positive measure such that for $a\in\AA$
equation \equ(nlw) has a solution
$$
u(x,t)=\sum_{i=1}^d a_i\cos(\mu_{n_i}'t+\theta_{i})\psi_{n_i}(x)+
\OO(|a|^3),
\equation(sasol)
$$
with frequencies $\mu'_{n_i}=\mu_{n_i}+\OO(|a|^2)$.
Furthermore, the set $\AA$ is of asymptotically full measure as
$|a|\rightarrow0$.

As is well known, the nonlinear wave equation \equ(nlw)
can be studied as an infinite
dimensional Hamiltonian system by taking the phase space to be the
product of the Sobolev spaces
$H^1_0([0,2\pi])\times L^2([0,2\pi])$ with
coordinates $u$ and $v=\partial_tu$.
The Hamiltonian for \equ(nlw) is then
$$
H=\onehalf(v,v)+\onehalf(Lu,u)+\int_0^{2\pi} g(u)\,dx,
\equation(hamilt1)
$$
where $L=-d^2/dx^2+m$, $g=\int fds$, and $(\cdot,\cdot)$ denotes the
usual scalar product in $L^2([0,2\pi])$.
In order to prove existence of solutions of the type \equ(sasol) by
means of \clm(mainthm), we would like to
write \equ(hamilt1) in the form \equ(hamiltonian).
This turns out to be possible, through amplitude-frequency modulation,
due to the availability of a (partial) normal
form theory for \equ(hamilt1). As we shall see, the requirement for
the parameter $m$ to be non zero is crucial for
this part of the argument.
In the sequel, we will closely follow the exposition
of P\"oschel in [P2].
Introducing the coordinates $\rq=(\rq_0,\rq_1,\rq_{-1},\dots)$ and
$\rp=(\rp_0,\rp_1,\rp_{-1},\dots)$ by setting
$$
u(x)=\sum_{n\in\integer}\rq_n\psi_n(x),\quad
v(x)=\sum_{n\in\integer}\rp_n\psi_n(x),
\equation(decompuv)
$$
one rewrites the Hamiltonian \equ(hamilt1) in
the coordinates $(\rq,\rp)$,
$$
H={1\over2}\sum_{n\in\integer}\bigl(\mu_n^2\rq_n^2+\rp_n^2\bigr)
+G(\rq),
\equation(hamilt2)
$$
where
$$
G(\rq)=\int_{0}^{2\pi} g\Bigl(\sum_{n\in\integer}\rq_n
\psi_n(x)\Bigr)dx.
\equation(nonlin)
$$
The Hamiltonian flow generated by \equ(hamilt2) is given by the
equations of motion
$$
\ddot \rq_n=-\mu_n^2\rq_n-\partial_{\rq_n}G(\rq),
\equation(eqmot4)
$$
and one can show that a solution $\rq$
of \equ(eqmot4) yields a solution of
the nonlinear wave equation \equ(nlw) if $\rq$
has some decaying properties.
More precisely, defining $l_b^s$ to be
the Banach space of
all real valued bi-infinite sequences $w=(w_0,w_{1},w_{-1},\dots)$
with norm
$$
||w||_s=\sum_{n\in\integer}[n]^s|w_n|,
$$
where $[n]=\max(1,|n|)$, one has the
\claim Lemma(solnlw)
Let $s\geq2$. If a curve $I\rightarrow l_b^s$,
$t\mapsto\rq(t)$,
is a solution of \equ(eqmot4),
then
$$
u(x,t)=\sum_{n\in\integer}\rq_n(t)\psi_n(x)
$$
is a classical solution of \equ(nlw).

For the proof of \clm(solnlw), see [CY].
Before turning to the normal form analysis of the Hamiltonian
\equ(hamilt2), we state a result concerning the regularity of the
gradient $\partial_{\rq} G$.
\claim Lemma(regGq)
For all $s>0$, the gradient $\partial_\rq G$
is real analytic as a map from some
neighborhood of the origin in $l_b^s$ into $l_b^s$, with
$$
||\partial_\rq G(\rq)||_s=\OO(||\rq||_s^3).
\equation(boundgrad)
$$

\proof
We first note that $l_b^s$ is a Banach algebra with respect to
convolution of sequences, with
$$
||q*p||_s\leq\sum_{i,j\in\integer}[i]^s|q_{j-i}||p_j|
\leq\sup_{i,j\in\integer}\Bigl({[i]\over[j-i][j]}\Bigr)^s
||q||_s||p||_s
\leq 2^s||q||_s||p||_s.
\equation(Balgebra)
$$
Therefore, using the analyticity of $f(u)=u^3+\OO(u^4)$, one computes
that in a sufficiently small neighborhood of the origin,
$$
||f(u)||_s\leq C||\rq||_s^3.
\equation(int111)
$$
On the other hand, since
$$
\partial_{\rq_n}G(\rq)=\int_0^{2\pi}f(u)\psi_n(x)dx,
$$
the components of $\partial_\rq G(\rq)$ are the Fourier components of
$f(u)$ and \equ(boundgrad) follows from the estimate \equ(int111).
The regularity of $\partial_\rq G$ follows from the regularity of its
components and its local boundedness, cf. [PT] p. 138.
\qed

We now turn to the normal form analysis of \equ(hamilt2).
First, since $g(u)=\onefourth u^4+\OO(u^5)$, we find that
$$
G(\rq)={1\over4}\sum_{i,j,k,l}g_{ijkl}\,\rq_i\rq_j\rq_k\rq_l
+\OO(|\rq|^5),
$$
where
$$
g_{ijkl}=\int_{0}^{2\pi}\psi_i\psi_j\psi_k\psi_ldx.
\equation(Nijkl)
$$
An easy computation shows
that $g_{ijkl}=0$ unless $i\pm j\pm k\pm l=0$ for at
least one combination of plus and minus signs.
This will play an important role later on.
Next, given a finite subset of indices
$\II_d=\{n_1,\dots,n_d\}\subset\integer$
with $|n_i|\not=|n_j|$ if $i\not=j$,
we decompose the
Hamiltonian \equ(hamilt2) as
$$
H=H_d+H_\infty,
$$
where
$$\eqalignno{
H_d(\rq,\rp)&={1\over2}\sum_{n\in\II_d}(\mu_n^2\rq_n^2+\rp_n^2)+
{1\over4}\sum_{i,j,k,l\in\II_d}g_{ijkl}\,\rq_i\rq_j\rq_k\rq_l
\equiv\Lambda_d(\rq,\rp)+G_d(\rq),\quad\qquad&\equat(Hd)\cr
H_\infty(\rq,\rp)&={1\over2}\sum_{n\not\in\II_d}(\mu_n^2\rq_n^2
+\rp_n^2)+
G(\rq)-G_d(\rq)\equiv\Lambda_\infty(\rq,\rp)
+G_\infty(\rq).&\equat(Hinfty)\cr
}$$
Introducing the complex coordinates $z_j$, $j=1,\dots,d$, by
$$
z_j={1\over\sqrt{2\mu_{n_j}}}(\mu_{n_j}\rq_{n_j}
+\imun\,\rp_{n_j})\,,
$$
one obtains the Hamiltonian
$H_d(z,\bar z)=\sum_j\mu_{n_j}|z_j|^2+G_d(z,\bar z)$ on
$\complex^d$ with
symplectic structure $\imun\sum_jdz_j\wedge d\bar z_j$.
For the remaining coordinates, one introduces the notation,
for $k\geq1$,
$$
x_k=\cases{(\rq_k,\rq_{-k})\in\real^2& if\quad
$k,-k\not\in\II_d\,$,\cr
            \rq_{-\tilde k}\in\real       & if\quad $k=|\tilde k|$
for some $\tilde k\in\II_d\,$,\cr}
$$
and similarly for $\rp_n$, $n\not\in\II_d$, denoted in terms of
$y_k\in\real^{d_k}$, $k\geq1$, with $d_k$ as above, namely, $d_k=2$ if
both $k,-k\not\in\II_d$ and $d_k=1$ otherwise.
Clearly, for $\rq,\rp\in l_b^s$ one has $x,y\in\RR^\infty_s$,
where $\RR^\infty_s$ is defined in \equ(defRRs),
and $H_\infty$ reads in these notations
$$
H_\infty(z,\bar z,x,y)={1\over2}\sum_{k\geq1}(\mu_k^2|x_k|^2+|y_k|^2)
+G_\infty(z,\bar z,x),
$$
with $|G_\infty|=\OO\bigl(\sum_{l=0}^3|z|^l||x||_s^{4-l}\bigl)$.
The next proposition establishes the existence of a symplectic change
of coordinates that transforms the Hamiltonian $H_d$ into a Birkhoff
normal form.
As it will be clear from the proof, this normal form is {\sl not}
available for
$H=H_d+H_\infty$, since most frequencies in $H_\infty$ are
degenerate.
This is the main difference with [P2] in the present discussion.

\claim Proposition(normalform)
For each $m>0$ and each subset $\II_d\,$, $d<\infty$, satisfying
$|n_i|\not=|n_j|$ when
$i\not=j$, there exists a near identity, real analytic, symplectic
change of coordinates $\Gamma_d$ in some neighborhood
of the origin in
$\complex^d$ that takes the Hamiltonian \equ(Hd) into
$$
H_d\circ\Gamma_d=\Lambda_d+\bar G_d+K_d,
$$
where $|K_d|=\OO(|z|^5)$ and
$$
\bar G_d(z,\bar z)={1\over2}\sum_{i,j=1}^d\bar g_{ij}|z_i|^2|z_j|^2
\quad{\rm with}\quad
\bar g_{ij}={3\over\pi}\,{4-\delta_{ij}\over\mu_{n_i}\mu_{n_j}}\,.
\equation(Gbar)
$$
Furthermore, setting
$\Gamma_\infty=\Gamma_d\oplus
\identity_{\RR^\infty_s\times\RR^\infty_s}$,
one has
$H_\infty\circ\Gamma_\infty=\Lambda_\infty+K_\infty$
with $|K_\infty|=\OO\bigl(\sum_{l=0}^3|z|^l||x||_s^{4-l}\bigr)$.

\proof
Modulo straightforward modifications, the proof is carried out in [P2]
and we restrict ourselves here to a quick overview.
The possibility to eliminate all terms in $G_d(z,\bar z)$
that are not of
the form $|z_i|^2|z_j|^2$
follows from the fact that for integers $i,j,k,l\in\II_d$ satisfying
$i\pm j\pm k\pm l=0$ {\rm and} $\{i,j,k,l\}\not=\{n,n,n',n'\}$
one has, as shown in [P2],
$$
|\mu_i\pm\mu_j\pm\mu_k\pm\mu_l|
\geq c\,{m\over(N^2+m)^{3/2}}>0\,,
\equation(crucial)
$$
with $c$ some absolute constant and $N=\min\{|i|,\dots,|l|\}$.
To see this, it is convenient to adopt the notation $z_j=w_j$ and
$\bar z_j=w_{-j}$ in which $G_d$ reads
$$
G_d=\sump_{i,j,k,l}\tilde g_{ijkl}w_iw_jw_kw_l+\OO(|z|^5),
\quad\tilde g_{ijkl}={g_{n_{|i|}\dots n_{|l|}}\over
\sqrt{\mu_{n_{|i|}}\dots\mu_{n_{|l|}}}}\ ,
$$
where the prime symbol in the summation sign indicates that the sum
runs over all indices $i,j,k,l\in\{1,-1,\dots,d,-d\}$ with
$n_{|i|}\pm n_{|j|}\pm n_{|k|}\pm n_{|l|}=0$
for at least one combination of plus and
minus signs.
Defining the transformation $\Gamma_d$ as the time-$1$ map of the flow
of the vector field $X_F$ given by a Hamiltonian $F(z,\bar z)$
of order
four, namely, $\Gamma_d=X^t_F|_{t=1}$ and
$F=\sump F_{ijkl}w_iw_jw_kw_l$, one obtains using Taylor's
formula $H_d\circ\Gamma_d=\Lambda_d+G_d+\{\Lambda_d,F\}+\OO(|z|^6)$
with
$$
\{\Lambda_d,F\}=-\imun\sump_{i,j,k,l}
(\hat\mu_{i}+\hat\mu_{j}+\hat\mu_{k}+\hat\mu_{l})
F_{ijkl}w_iw_jw_kw_l,
$$
where $\hat\mu_i\equiv\sign(i)\mu_{n_{|i|}}$.
Therefore, \equ(crucial) allows to choose $F_{ijkl}$ in such a way
that
$$
G_d+\{\Lambda_d,F\}=\sum_{i,j=1}^d\tilde g_{iijj}|z_i|^2|z_j|^2
+\OO(|z|^5)\equiv
\bar G_d+\OO(|z|^5)\,.
$$
For the rest of the proof, we refer the reader to [P2].
\qed
The Hamiltonian $\Lambda_d+\bar G_d$ is integrable with
integrals $|z_i|^2, i=1,\dots,d$. Furthermore, the matrix
$\bar g=(\bar g_{ij})_{i,j}$ is non degenerate, as can be
checked from the explicit
formula \equ(Gbar). Hence, introducing the standard action-angle
variables $(I,\phi)\in\real^d\times\torus^d$ and linearizing $H$
around a given value for the action,
namely, by setting
for some $a=(a_1,\dots,a_d)\in\real^d$,
$$
z_i\bar z_i=I_i+a_i^2,
$$
one finally obtains
$$
H_a=\omega\cdot I+\onehalf I\cdot\bar gI
+\sum_{k\geq1}(\mu_k^2x_k^2+y_k^2)+U_a(I,\phi,x),
\equation(Hfinal)
$$
where $U_a$ is just $K_d+K_\infty$
with the variables $z_i,\bar z_i, i=1,\dots,d$, expressed in terms of
$I,\phi$, and where $\omega=(\omega_1,\dots,\omega_d)$ is given by
$$
\omega_i=\mu_{n_i}+\sum_{j=1}^d\bar g_{ij}a_j^2\,,
$$
and covers a cone at $(\mu_{n_1},\dots,\mu_{n_d})$ as $a$
varies in a neighborhood of the origin of $\real^d$.
Furthermore, $U_a$ is real analytic in
$\phi\in\torus^d$ and real analytic in $I$ in a sufficiently small
neighborhood $\sO_I$ of the origin of $\real^d$.
As a function of $x$, $U_a$ is real
analytic in a neighborhood $\sO_x\subset\RR_s^\infty$
and by \clm(regGq), its gradient
$\partial_x U_a$ is bounded as a map from
$\torus^d\times\sO_I\times\sO_x$ to $\RR^\infty_s$.
Therefore, since hypothesis (H1) is satisfied with $\gamma=1$,
$U_a$ satisfies (H2) with $\xi=1$.
Finally, the small parameter $\lambda$ is given in terms of
$|a|=\delta$.
In the Hamilton's equations for $H_a$, rescaling
$a$ by $\delta$,
$x$ and $y$ by $\delta^2$, and $I$ by $\delta^4$,
one obtains an Hamiltonian system given by the
rescaled Hamiltonian
$$\eqalign{
\tilde H_a(\phi,I,x,y)&=
\delta^{-4}H_{\delta a}(\phi,\delta^4I,\delta^2 x,\delta^2 y)\cr
&=\omega\cdot I+{\delta^4\over2}I\cdot\bar gI
+\sum_{k\geq1}(\mu_k^2x_k^2+y_k^2)+
\tilde U_a(I,\phi,x),
}$$
with $\tilde U_a$ analytic in $\delta$ and,
as a function of $I$,
$$
\tilde U_a=\OO(\delta)+\OO(\delta^3|I|)+\OO(\delta^5|I|^2).
$$
Hence, \clm(mainthm) implies the existence of quasi-periodic solutions
$I,x$ and $y$ of period $\omega$, real analytic in $\phi$ and
$\lambda$. Tracing the coordinate transformations back to the original
variables $\rq_n(t)$ in the expression \equ(decompuv) for $u(x,t)$
completes the proof of \clm(thmnlw) with $u(x,t)$ given
by \equ(sasol).

\section The Renormalization Group Scheme

Equation \equ(eqmot3) consists in a system of
equations for the variables $(\Phi,J)$ and $Z$ which
are coupled through the perturbation $U$ only.
Adopting the notation
$$\eqalignno{
V(\Phi,J,Z)(\varphi)&=
\lambda\pmatrix{\partial_\phi U\cr
                \partial_I U\cr}
(\varphi+\Phi(\varphi),J(\varphi),Z(\varphi)),
&\equat(defV)\cr
W(\Phi,J,Z)(\varphi)&=
\lambda\partial_xU(\varphi+\Phi(\varphi),J(\varphi),Z(\varphi)),
&\equat(defW)\cr
}$$
one rewrites equation \equ(eqmot3) as
$$\eqalignno{
\pmatrix{0&D\cr-D&g}\pmatrix{\Phi\cr J}&=-V(\Phi,J,Z),&\equat(sys1)\cr
(D^2+\mu^2)Z&=-W(\Phi,J,Z).&\equat(sys2)\cr
}$$
Our strategy will be
to consider \equ(sys1) and \equ(sys2) separately, treating the
functions $Z$ and $(\Phi,J)$, respectively, as parameters.  As we will
see in Section~8, existence of a (unique) solution of the original
equation \equ(eqmot3) can then be proved by using the implicit
function theorem.  Note that \equ(sys1) involves only the torus
frequencies $\omega$ and is equivalent to a standard KAM
problem. Existence of solution for such equations is well known and
has been established by various means. One important feature we will
use is the regular dependence of the solution $(\Phi,J)$ on the
function $Z$. A precise result about the solution of \equ(sys1) will
be stated in Section~4, Theorem~4.1, once the required Banach spaces
of functions have been introduced.

We now focus our attention on equation \equ(sys2), and will
suppress from the notation the dependence of the vector field
$W$ on the
parameters $\Phi$ and $J$. Most of our analysis will be conducted in
Fourier space, and we will denote by lower case letters
the Fourier transforms
of functions of $\varphi$, the latter being denoted by capital
letters, namely,
$$
F(\varphi)=\sum_{q\in\integer^d}
e^{-iq\cdot\varphi}f(q),\quad
{\rm where}\quad f(q)=\int_{\torus^d}e^{iq\cdot\varphi}
F(\varphi)d\varphi,
$$
where $d\varphi$ stands for the normalized Lebesgue measure on
$\torus^d$.
For $Z(\varphi)\in\RR^\infty$, note that $z(q)\in\hat\RR^\infty$ with
$z_{k_i}(q)=\overline{z_{k_i}(-q)}$, where $\hat\RR^\infty$ stands for
$\bigoplus_{k\geq1}\complex^{d_k}$
and $k_i$ refers to the $i^{\rm th}$ component of $\complex^{d_k}$.
Similarly,
$\hat\RR^\infty_s$ will denote the
complexification of the Banach space $\RR^\infty_s$
defined in \equ(defRRs). Finally, we will denote the vector
space of functions
$z(q)\in\hat\RR^\infty$ by $h$,
$$
h=\{z=(z(q))\,|\,z(q)\in\hat\RR^\infty,q\in\integer^d\}.
$$
In terms of the Fourier transform of $W$, namely,
$$
w_0(z)(q)\equiv\lambda\int_{\torus^d}e^{iq\cdot\varphi}\partial_x
U(\varphi+\Phi(\varphi),J(\varphi),Z(\varphi))d\varphi,
\equation(wrho)
$$
equation \equ(sys2) becomes,
$$
\sK_0z=w_0(z),
\equation(fp1)
$$
where the operator $\sK_0$ is given by the diagonal kernel
$$
\sK_0(q,q')=\bigl(|\omega\cdot q|^{2}-\mu^2\bigr)
\delta_{qq'}.\equation(drho)
$$
Solving equation \equ(fp1) requires to invert the operator $\sK_0$.
Although the inverse of $\sK_0$ is unbounded for generic frequencies,
restricting $\omega$ to a set of admissible frequencies gives
sufficient control on the inverse of $\sK_0$ to prove existence of a
solution. As is well known for Melnikov problems, this set
depends on the perturbation $U$.

In order to prove existence of a solution to equation \equ(fp1), we
will follow a strategy developed in [BGK] for standard KAM problems,
namely, for equations of the type \equ(sys1).
This strategy basically consists in
inductively reducing \equ(fp1) to a sequence of
effective equations involving denominators of decreasing size.
One inductive step, say
the $n^{\rm th}$ step, consists in splitting the effective equation
obtained at the previous step
into two equations involving only large and, respectively, small
denominators, where large and small are defined with respect to a
scale of order $\eta^n$ for some fixed $\eta<1$. This
splitting is done in such a way that the nonlinear operator involved
in the large denominators equation is a contraction, and this equation
can thus be solved by a simple application of the contraction mapping
principle. This, in turn, allows to map the small
denominators equation
into a new effective equation of the type \equ(fp1), with a new right
hand side $w_n$ and (eventually) a new linear operator $\sK_n$.
In [BGK], it was shown that for equations of the type \equ(sys1),
the above mentioned contraction
property follows naturally from symmetries
specific to this case.
In contrast, equation \equ(sys2)
involves in addition the normal
frequencies $\mu_k$ and does not possess such symmetry.
In order to obtain
the required contraction, we must make at every inductive step an
additional preparation step.
As we shall see below, this amounts to renormalizing
the linear operator
$\sK_{n-1}$ obtained at the previous step into a new operator $\sK_n$,
which, in effect, corresponds to renormalizing the
normal frequencies. Furthermore, we will see that
the renormalized normal frequencies converge
to a $U$-dependent set $\{\mu_\alpha^*\},\ \alpha\geq1,$ as
$n\rightarrow\infty$. Therefore,
since the Diophantine conditions
imposed on $\omega$ will eventually be defined relatively to this set,
one obtains in a
constructive way the dependence of the set of
admissible frequencies on the perturbation $U$.

We now describe how the
renormalization group approach is implemented
in practice for Melnikov type problems.
First, we proceed with the above mentioned preparation step
by decomposing $w_0$ as
$$
w_0(z)=\tilde w_0(z)+A_0z,
$$
where the linear operator $A_0$ is the dominant part of
$Dw_0(z)$ evaluated at $z=0$.
With $\sK_1\equiv \sK_0-A_0$, equation \equ(fp1) now reads
$$
\sK_1z=\tilde w_0(z).
\equation(step1)
$$
As explained in more details below, $A_0$ can be chosen in such a way
that $\sK_1$ is of the same form as $\sK_0$, cf. \equ(drho),
but now given
in terms of a new set of frequencies $\tilde\mu_{k_i}\in\real$
which are
perturbation of order $\lambda$ of the original normal frequencies
$\mu_k$. The notation $\tilde\mu_{k_i}$ reflects the fact that
the perturbation $A_0$ may lift some of the degeneracies.
Therefore, when inverting $\sK_1$, denominators smaller than
$\OO(\eta)$
occur for $q$ such that
$||\omega\cdot q|-\tilde\mu_{k_i}|\leq\OO(\eta)$
for some $k_i$. Furthermore, these small denominators only occur,
for such $q$, in a specific
subspace $h_{k_i}^q$ of $\complex^{d_k}$ depending on which
$\tilde\mu_{k_j}$, if any, has been separated from
$\tilde\mu_{k_i}$ by more
than $\OO(\eta)$.
Introducing $P_1$ as the projection of $h$
onto $h^q_{k_i}$ for $q$ such that
$||\omega\cdot q|-\tilde\mu_{k_i}|\leq\OO(\eta)$ and defining
$Q_1\equiv\identity-P_1$, one thus expects that
the restriction of $\sK_1$
to $Q_1h$ is invertible with an inverse of order $\OO(\eta^{-1})$.
Multiplying \equ(step1) by $Q_1$ and $P_1$ leads to the
small and large denominators equations
for $\tilde z_1\equiv Q_1z$ and $z_1\equiv P_1 z$,
$$\eqalignno{
\sK_1\tilde z_1&=Q_1\tilde w_0(\tilde z_1+z_1),&\equat(step11)\cr
\sK_1z_1&=P_1\tilde w_0(\tilde z_1+z_1),&\equat(step12)\cr
}$$
and by definition of $Q_1$,
the first equation
can be rewritten as a fixed point equation for the functional $R_1$
defined as $R_1(z_1)\equiv\tilde z_1$, namely,
$$
R_1(z)=\sK_1^{-1}Q_1\tilde w_0(z+R_1(z)).
\equation(step121)
$$
By choice of $A_0$, the nonlinear operator
$\sK_1^{-1}Q_1\tilde w_0$ is
a contraction and one can solve equation \equ(step121) for
$R_1$ using the
Banach fixed point theorem.
(See point (a) of Theorem~5.1 for this part
of the inductive step.)
Next, with $w_1$ defined as
$$
w_1(z)\equiv\tilde w_0(z+R_1(z)),
$$
equation \equ(step12) reads
$$
\sK_1 z_1=P_1 w_1(z_1),
\equation(step13)
$$
and the solution $z=z_1+\tilde z_1$ of the
original equation \equ(fp1) is now given  by
$$
z=z_1+R_1(z_1)\equiv F_1(z_1).
$$
Hence, the problem of solving \equ(fp1) is reduced to solving
the effective equation \equ(step13).
To solve this equation one proceeds similarly, starting with our
preparation step.
After $n$ steps of this inductive process, the solution of \equ(fp1)
is given by
$$
z=F_{n-1}(z_n+R_n(z_n))\equiv F_{n}(z_n),
\equation(deffn)
$$
where
$R_n$ solves the functional equation
$$
R_n(z)=\Gamma_n\tilde w_{n-1}(z+R_n(z)),
\equation(equrn)
$$
with
$$
\Gamma_n\equiv \sK_{n}^{-1}Q_{n}P_{n-1},
\equation(defGamman)
$$
and, for some linear operator $A_{n-1}$,
$$\eqalignno{
&\tilde w_{n-1}(z)\equiv w_{n-1}(z)-A_{n-1}z,&\equat(defwtilden)\cr
&\sK_{n}\equiv \sK_{n-1}-P_{n-1}A_{n-1},&\equat(defAn)\cr
}$$
whereas $z_n$ solves the effective equation
$$
\sK_{n}z_n=P_n w_{n}(z_n),
\equation(stepn3)
$$
with $w_{n}$ defined as
$$
w_{n}(z)\equiv\tilde w_{n-1}(z+R_n(z)).
\equation(defwn)
$$
\claimrm Remark(rem0)
The point of this inductive procedure is that
$P_nw_n(z)$ becomes effectively linear in $z$ for large $n$.
More precisely, we will show, cf. Theorem~5.1 below, that the rescaled
maps $w_n^r$ defined by $w_n^r(z)=\eta^{-n}r^{-n}w_n(r^{n}z)$
satisfy for $r<\eta$,
$$
P_nw_n^r(z)=P_nDw_n^r(0)z+\OO(\lambda r^{2n}\eta^{-n})
\quad{\rm with}\quad P_nDw_n^r(0)=\OO(\lambda),
$$
in some appropriate Banach space.
Thus,
$z_n=0$ becomes a better and better approximation to the solution of
\equ(stepn3), and we shall construct the solution $z$ of the original
equation \equ(fp1) as the limit of the approximate solutions
$$
z=\lim_{n\rightarrow\infty}F_n(0).
\equation(zlimit)
$$

We now give a precise description of the operators $P_n$.
Note that in order
to obtain \equ(equrn) and \equ(stepn3), we have tacitly assumed
that $P_{n}P_{n-1}=P_n$. The possibility to define $P_n$
satisfying such a property follows from the convergence of the
normal frequencies under renormalization.
Recall that renormalization occurs because at every inductive step one
turns the nonlinear map $w_n$ of the effective functional equation
\equ(stepn3) into a
contraction by substracting some linear operator $A_n$. Delaying to
subsequent sections the
discussion of the appropriate choice for the family $A_m$, $m\geq0$,
it suffices to point here to the properties of $A_m$ that will ensure
convergence of the renormalized normal frequencies. As will be shown,
cf. point (c) of Theorem~5.1 for a precise statement,
$A_m$ is a perturbation of order
$\lambda\eta^{m}$ and is
given by a constant kernel $A_m(q,q')=a_m\delta_{qq'}$ with
$a_m:\hat\RR^\infty\rightarrow\hat\RR^\infty$ linear and hermitian.
As a consequence, the operator
$\sK_n=\sK_0-\sum_{m=0}^{n-1}P_mA_m$ has
a kernel of the form \equ(drho) with $\mu^2$ essentially replaced by
the positive definite matrix
$$
\tilde\mu_n^2\equiv\mu^2+\sum_{m=0}^{n-1}a_m,
\equation(defmun)
$$
with $\tilde\mu_n$ having a discrete spectrum
$\sigma(\tilde\mu_n)\subset\real^+$. One easily checks that the
singularities of $\sK_n^{-1}$ are given by the eigenvalues of
$\tilde\mu_n$, which therefore correspond to renormalized normal
frequencies. Since $a_m$ is of order
$\lambda\eta^{m}$, one expects the
eigenvalues of $\tilde\mu_n$ to converge as $n\rightarrow\infty$ with
$|\nu_{n+1}-\nu_{n}|\leq\OO(\lambda\eta^n)$ for
$\nu_{n+1}\in\sigma(\tilde\mu_{n+1})$ and
$\nu_{n}\in\sigma(\tilde\mu_{n})$.
This, in turn, allows us to define scales of denominators in a
consistent way by carefully keeping track of
the separation properties of
$\sigma(\tilde\mu_n)$ as $n$ increases. To this end, one
groups the normal frequencies into a
hierarchy of clusters satisfying gap conditions that are
preserved by the renormalization procedure.
We first introduce some notation.
For $x\in\real$ and
$\CC$ a finite collection of points in $\real$, let
$d(x,\CC)$ denote
the distance between $x$ and the smallest interval
containing all points in $\CC$, and
for two finite collections $\CC_1,\CC_2\subset\real$, let
$$
d(\CC_1,\CC_2)\equiv\inf_{x\in\CC_1}d(x,\CC_2).
$$
Then, one can uniquely decompose $\sigma(\tilde\mu_n)$ into
a maximal number of disjoint clusters $\CC^n_{k,i}$, $k\geq1$,
$i=1,\dots,M_k^n$,
satisfying $d(\mu_k,\CC^n_{k,i})=\OO(\lambda)$
and the gap condition
$$
d(\CC^n_{k,i},\CC^n_{k,j})>\eta^n\quad{\rm if}\ i\not=j.
\equation(gapn)
$$
Note that $M^n_{k}\leq d_k$, where $d_k$ denotes the multiplicity of
the original normal frequency $\mu_k$, and that
by requiring $M^n_{k}$ to be maximal, the decomposition
$$
\sigma(\tilde\mu_{n})=\bigcup_{k\geq1}
\bigcup_{i=1}^{M^n_{k}}\CC^n_{k,i}
\equation(defcin)
$$
is unique. The above observation about the rate of convergence of
$\sigma(\tilde\mu_n)$ as $n\rightarrow\infty$ ensures that eigenvalues
belonging to different clusters will remain separated.
Generically, one expects all degeneracies to be lifted eventually,
so that $M^n_k=d_k$ for $n$ sufficiently large and each cluster
$\CC^n_{k,i}$ contains a single eigenvalue.
Next,
defining $S_{n}\subset\integer^d$ as
$$
S_{n}=\bigcup_{k\geq1}\bigcup_{i=1}^{M^n_{k}}S^n_{k,i},
\equation(defsn)
$$
where
$$
S^n_{k,i}=\{q\in\integer^d\ |\
d(|\omega\cdot q|,\CC^n_{k,i})<\onefourth\eta^{n}\},
\equation(defsin)
$$
one is ensured that all $q\in\integer^d\setminus S_n$ satisfy
$d(|\omega\cdot q|,\sigma(\tilde\mu_{n'}))\geq\OO(\eta^n)$ for
$n'\geq n$. Hence, such $q$
can be safely ``integrated out'' in the large denominators
equation.
Remark that due to \equ(gapn), the sets $S^n_{k,i}$
are pairwise disjoint.
In order to achieve the construction of $P_n$, one must isolate for
every $q\in S_n$ the
subspace of $\RR^\infty$ in which small denominators will occur.
For $q\in S^n_{k,i}$, the latter is given by the eigenspace of
$\tilde\mu_n$ associated with the
eigenvalues belonging to $\CC^n_{k,i}$.
This eigenspace will be denoted
by $\JJ^n_{k,i}$, whereas the projector onto $\JJ^n_{k,i}$ will be
denoted by $\PP^n_{k,i}$.
Thus, one defines $P_n$ to be the diagonal operator acting on $h$
given by the kernel
$$
P_n(q)=\sum_{k\geq1}
\sum_{i=1}^{M^n_k}\chi^n_{k,i}(\omega\cdot q)\PP^n_{k,i},
\equation(defPn)
$$
where
$\chi^n_{k,i}$ denotes a function in
$\in\CC^1(\real)$ which satisfies
$$
\chi^n_{k,i}(\kappa)=\cases{
1&if\ \ $d(|\kappa|,\CC^n_{k,i})\leq\oneeighth\eta^n$,\cr
0&if\ \ $d(|\kappa|,\CC^n_{k,i})\geq\onefourth\eta^n$,\cr
}
$$
and interpolates monotonically
between $0$ and $1$ otherwise, with
$$
\sup_{\kappa\in\real}
|{\chi^n_{k,i}}'(\kappa)|\leq C\eta^{-n},
\equation(condder)
$$
whereas $Q_n$ is defined as
$$
Q_n=\identity-P_n\,.
\equation(defQn)
$$
Note that $P_n$ and $Q_n$ are not projectors. The smooth functions
$\chi^n_{k,i}$ have been introduced in order to ensure the continuity
of the diagonal kernels $\Gamma_n(q,q)$, cf. the
discussion preceding Lemma~5.3 below. However,
we will make use later of the projector
$$
\hat P_n(q)=\sum_{k\geq1}
\sum_{i=1}^{M^n_k}{\rm I}_{S^n_{k,i}}(q)\PP^n_{k,i},
\equation(defPhatn)
$$
where ${\rm I}_\Sigma$ denotes the indicator function of a set
$\Sigma$.
Note that $P_n\hat P_n=P_n$, whereas $Q_n\hat P_n\not=0$.

We conclude this section
by a few remarks related to
the convergence of the inductive scheme.
First, setting $I^n_{k,i}\subset\real$ to be the smallest interval
covering $\CC^n_{k,i}$, one easily checks that
$|I^n_{k,i}|\leq(d_k-1)\eta^n$.
Hence, since the multiplicities of the normal frequencies $\mu_k$
were assumed to be uniformly bounded in $k$, {i.e.},
$d_k\leq\bar d$ for all $k\geq1$, one obtains
for all $n\geq1$, $k\geq1$, and $i=1,\dots,M^n_{k}$,
$$
|I^n_{k,i}|\leq\bar d\eta^n.
\equation(bndabsIin)
$$
Next, it follows from the gap condition \equ(gapn) being preserved
that for all $m<n$ the eigenvalues in a given cluster $\CC^n_{k,i}$
are perturbation of all or
some eigenvalues belonging to a single cluster $\CC^m_{k,j}$, denoted
by $\CC^m_{k,j^n_i}$. Furthermore, $\CC^n_{k,i}$ remains close to
$\CC^m_{k,j^n_i}$.
More precisely, we will show that
$$
\sup_{x\in I^n_{k,i}}\inf_{y\in I^m_{k,j^n_i}}d(x,y)\leq\eta^{m+1}
\quad{\rm for}\quad 1\leq m<n.
\equation(convclust)
$$
Finally, we consider the properties of the eigenspaces
$\JJ^n_{k,i}$. One has by construction
$\PP^n_{k,i}\PP^n_{l,j}=\delta_{kl}\delta_{ij}\PP^n_{k,i}$.
However, it will be possible to chose $a_{m}$ in \equ(defmun)
in such a way that
each $\JJ^{m}_{k,i}$ is an invariant subspace for $a_{m}$.
Hence, by definition of $\tilde\mu_n$ and
$\JJ^n_{k,i}$, every $\JJ^n_{k,i}$ is a subspace of some
$\JJ^{n-1}_{k,j}$, and by recursion, of some $\JJ^{m}_{k,j}$
for all $m<n$. The (unique) eigenspace $\JJ^{m}_{k,j}$ containing
$\JJ^n_{k,i}$ will be denoted by $\JJ^{m}_{k,j^n_i}$.
Therefore, one has
for all $1\leq m\leq n$, $k\geq1$, and $i=1,\dots,M^n_k$,
$$
\PP^n_{k,i}\PP^m_{l,j}=\delta_{kl}\delta_{jj^{n}_{i}}\PP^n_{k,i},
\equation(PkiPlj)
$$
which, in particular, implies that
$$
P_{n}P_{n-1}=P_{n-1}P_{n}=P_{n}.
\equation(proppn)
$$

\notation
For most of the subsequent analysis, it will not be necessary to
distinguish between indices $(k,i)$ and $(l,j)$ with $k=l$ or
$k\not=l$. This intervenes only in the description of the asymptotic
behavior of the spectrum $\sigma(\tilde\mu_n)$ and
the measure estimate
of $\Omega^*$. For notational convenience, we thus introduce the
index sets
$$
\II^n=\{(k,i)\ |\ k\geq1,\ i=1,\dots,M^n_k\},\quad n\geq1,
\equation(defSS)
$$
and will reserve bold letters for indices in $\II^n$.
With this convention, $\{\CC^n_{\bf k}\ |\ {\bf k}\in\II^n\}$,
denotes for instance the
collection of all clusters $\CC^n_{k,i}$, $k\geq1$, $i=1,\dots,M^n_k$.

\section Spaces

For the Fourier transform $z$ of the solution $Z$ of our
original equation \equ(sys2), we consider the Banach space $h_s$,
$s\in\real$, defined by
$$
h_s=\{z=(z(q))\ |\ z(q)\in\hat\RR^\infty_s,\
||z||_s\equiv\sum_{q\in\integer^d}|z(q)|_s<\infty\}.
\equation(spaceh)
$$
For $s\geq t$, one has the natural embedding
$h_s\rightarrow h_t$ with
$||\cdot||_t\leq||\cdot||_s$.
We will denote by $h^n_s$ the subspace $\hat P_nh_s$.
In particular, one has for $z\in h^n_s$,
$$
||z||_s=
\sum_{{\bf k}\in\II^n}\sum_{q\in S^n_{\bf k}}
|\PP^n_{\bf k}z(q)|_{s}.
\equation(normhn)
$$
The operator norm in $\LL(h^n_s,h^m_t)$ will be denoted
by $||\cdot||^{(n,m)}_{s,t}$, and by $||\cdot||^{(n)}_{s}$ when $n=m$
and $s=t$.

Let us now turn to the spaces we will consider
for the functions $w_n$.
Recall that in our analysis of \equ(sys2), the functions
$\Phi$ and $J$
only appear as parameters. In the sequel, we consider
$\Phi,J:\torus^d\rightarrow\real^d$ as (fixed) real analytic maps
belonging to a small neighborhood of the origin $\sO_\sB$ in the
Banach space
$$
\sB=\{(F,G):\torus^d\rightarrow\real^d\times\real^d\,|\,
||(F,G)||_{\sB}\equiv\sum_{q\in\integer^d}|f(q)|+|g(q)|<\infty\}.
\equation(defsB)
$$
Next, it follows from assumption (H2) that the gradient
$\partial_xU$ is
real analytic as a map from $\torus^d\times\sO_I\times\sO_x$ to
$\RR^\infty_{s'}$, cf. [PT] p. 138.
(Recall that
$\sO_I\subset\real^d$ and $\sO_x\subset\RR^\infty_s$
are neighborhoods of
the origin and that $s'\equiv s+\xi-a$.)
This implies that for $(\Phi,J)\in\sO_\sB$ small enough,
one can write the Taylor expansion of
$\partial_xU(\varphi+\Phi(\varphi),J(\varphi),Z)=
\partial_xU((\varphi+\Phi(\varphi),J(\varphi),0)+(0,0,Z))$
as
$$
\partial_xU((\varphi+\Phi(\varphi),J(\varphi),0)+(0,0,Z))=
\sum_{m=0}^\infty
{1\over m!}U_{m+1}^{\Phi,J}(\varphi)(Z,\dots,Z),
\equation(expa1)
$$
where the coefficients $U^{\Phi,J}_{m+1}(\varphi)$
belong to the space
of $m$-linear maps $\LL(\RR_s,\dots,\RR_s;\RR_{s'})$, are real
analytic in $\varphi\in\torus^d$ and analytic in
$(\Phi,J)\in\sO_\sB$.
Hence, there exist $\rho>0$, $\alpha>0$ and $b<\infty$
such that the Fourier
transforms of $U^{\Phi,J}_{m+1}(\varphi)$ satisfy
$$
\sum_{q\in\integer^d}e^{\alpha|q|}\,||u_{m+1}^{\Phi,J}(q)||_
{\LL(\hat\RR_s,\dots,\hat\RR_s;\hat\RR_{s'})}\leq b\,m!\,\rho^{-m}.
\equation(coeff1)
$$
Inserting the Fourier series for $Z$ into \equ(expa1), one obtains the
expansion for $w_0$ as defined in \equ(wrho),
$$\eqalignno{
w_0(z)(q)&=\lambda\sum_{m=0}^\infty\sum_{{\bf q}}{1\over m!}
u^{\Phi,J}_{m+1}\Bigl(q-\sum_{i=1}^mq_i\Bigr)(z(q_1),\dots,z(q_m))&\cr
&\equiv\sum_{m=0}^\infty\sum_{{\bf q}}
w^{(m)}_0(q;q_1,\dots,q_m)(z(q_1),\dots,z(q_m)),&\equat(expa2)\cr
}$$
where ${\bf q}=(q_1,\dots,q_m)\in\integer^{md}$.
This formula suggests to consider $w_0$ as an analytic functions
 of
$z\in h_s$.
Let $B(r_0)$ be the open ball of radius $r_0$ in $h_s$ centered at the
origin and let $H^\infty(B(r_0),h_{s'})$ denote the Banach space of
analytic function $w:B(r_0)\rightarrow H_{s'}$
equipped with the supremum
norm, which we shall denote by $|||w|||$.
Then, bound \equ(coeff1) implies that $w_0\in H^\infty(B(r_0),h_{s'})$
for $r_0$ small enough.

It will be convenient to encode the decay property of the kernels
$w^{(m)}_0$ inherited from the estimate \equ(coeff1) as a property of
the functional $w_0$. Let $\tau_\beta$ denote the
translation by $\beta\in\real^d$, i.e.,
$(\tau_\beta Z)(\varphi)=Z(\varphi-\beta)$.
On $h_s$, $\tau_\beta$ is realized by
$(\tau_\beta z)(q)=e^{i\beta\cdot q}z(q)$, and it induces a map
$w\mapsto w_\beta$ from $H^\infty(B(r_0),h_{s'})$ to itself if we
define
$$
w_\beta(z)=\tau_\beta w(\tau_{-\beta}z).
\equation(taubeta)
$$
On the kernels $w_0^{(m)}$, this is given by
$$
w_\beta^{(m)}(q;q_1,\dots,q_m)=e^{i\beta\cdot(q-\sum q_i)}
w^{(m)}(q;q_1,\dots,q_m),
$$
and makes sense also for $\beta\in\complex^d$.
Since
$$
|||w_{0\beta}|||\leq\sum_{m=0}^\infty r_0^m\sup_{\bf q}
\sum_{q\in\integer^d}e^{-{\rm Im}\beta\cdot((q-\sum q_i)}
||w_0^{(m)}(q;q_1,\dots,q_m)||_{\LL(\hat\RR_s,\dots,
\hat\RR_s;\hat\RR_{s'})},
$$
it thus follows from \equ(coeff1) that there exist
$r_0>0$, $\alpha>0$, and $D<\infty$,
such that $w_{0\beta}$ belongs to $H^\infty(B(r_0),h_{s'})$
and extends
to an analytic function of $\beta$ in the strip $|\Im\beta|<\alpha$
with values in $H^\infty(B(r_0),h_{s'})$ satisfying the bound
$$
|||w_{0\beta}|||\leq D|\lambda|.
\equation(nw0beta)
$$

Let us now come back to the existence of a solution for equation
\equ(sys1), namely for the standard KAM problem.
One has the classical
result (see for instance [BGK]):
\claim Theorem(KAM)
Let $U$ satisfy hypothesis (H2) and let $g$
be an invertible matrix. Then, there is a
$\lambda_1>0$ small enough such that
for $|\lambda|<\lambda_1$ and
$\omega$ satisfying a Diophantine condition of the form
$$
|\omega\cdot q|>K|q|^{-\nu}\quad{\rm for}\quad q\in\integer^d,
\,q\not=0,
$$
\equ(sys1) has a solution $(\Phi,J)\in\sB$
which is real analytic in $\varphi$,
analytic in $\lambda$, and vanishes for $\lambda=0$.
Furthermore, this solution is unique up to translations
$(\Phi,J)(\varphi)\mapsto(\Phi-\beta,J)(\varphi-\beta)$
and depends analytically on $Z$, for $Z$ in a small ball centered at
the origin of the Banach space $h_s$.

To conclude this section,
we list some standard
properties of bounded
analytic functions defined on open balls in Banach spaces. Let
$h,h',h''$ be Banach spaces, $B(r)\subset h$, $B(r')\subset h'$, and
$w_i\in H^\infty(B(r),h')$, $w\in H^\infty(B(r'),h'')$. First, one has
the composition property: If $|||w_i|||<r'$ then
$w\circ w_i\in H^\infty(B(r),h'')$ and
$$
|||w\circ w_i|||<|||w|||.
\equation(comp)
$$
Next, one deduces from the Cauchy estimate that for $r_1<r'$,
$$
\sup_{||x||<r_1}||Dw(x)||_{\LL(h',h'')}\leq(r'-r_1)^{-1}|||w|||.
\equation(Cauchy1)
$$
Taking $r_1=\onehalf r'$, we infer that if $|||w_i|||\leq\onehalf r'$
then
$$
|||w\circ w_1-w\circ w_2|||\leq{2\over r'}|||w|||\,|||w_1-w_2|||.
\equation(Cauchy2)
$$
Moreover, if
$\delta_kw(x)\equiv w(x)-\sum_{l=0}^{k-1}{1\over l!}D^lw(0)(x)$,
then
$$
\sup_{||x||\leq\gamma r'}||\delta_kw(x)||
\leq{\gamma^k\over1-\gamma}|||w|||,
\equation(Cauchy3)
$$
for $0\leq\gamma<1$.

\section Inductive Bounds

We now turn to the inductive bounds that will be used
to prove \clm(mainthm).
We first note that since in \equ(equrn) and \equ(defwtilden),
$\Gamma_n$ and $A_n$ are diagonal operators,
applying $\tau_\beta$ to equation \equ(equrn) leads to
$$
R_{n\beta}(z)=\Gamma_n\tilde w_{(n-1)\beta}(z+R_{n\beta}(z)),
\equation(eqrnb)
$$
where $\tilde w_{(n-1)\beta}=w_{(n-1)\beta}-A_{n-1}$, and
$w_{n\beta}$ is now recursively defined by
$$
w_{n\beta}(z)\equiv\tilde w_{(n-1)\beta}(z+R_{n\beta}(z)).
\equation(defwnb)
$$

For $r<1$ a parameter to be chosen later, let $B_n$
denote the open
ball of radius $r^{n+1}$ in $h^n_s$ centered at the origin.
Then, we will show that $w_{n\beta}$ belongs to
$H^\infty(B_n,h_{s'})$,
the Banach space of analytic functions $w:B_n\rightarrow h_{s'}$,
provided
$|\lambda|$ is taken small enough (uniformly in $n$) and provided
the analyticity strip in $\beta$ is restricted slightly.
In the sequel, we will denote $H^\infty(B_n,h_{s'})$ by $\AA_n$.
As mentioned in \clm(rem0),
the main ingredient in proving \clm(mainthm) is to show that
in addition, $\hat P_nw_{n\beta}$ becomes essentially linear as
$n\rightarrow\infty$.
Before stating this result,
one introduces the following
frequency subsets, setting
for $K>0$ and $\{\CC^n_{\bf k}\}_{{\bf k}\in\II^n}$ the
clusters described in the previous section,
$$\displaylines{\quad
\Omega_n(K)=\bigl\{\omega\in\real^d\ |\
\ d(|\omega\cdot q|,\CC^n_{\bf k}),\
d(|\omega\cdot q|,|\CC^n_{\bf k}\pm\CC^n_{\bf k'}|)
>K|q|^{-\nu}\hfill\cr
\hfill\forall\,|q|<K\eta^{-n/\nu}, q\not=0,\
{\rm and}\ {\bf k},{\bf k}'\in\II^n\},\quad
\equat(omegank)\cr
}$$
where $\CC^n_{\bf k}\pm\CC^n_{\bf k'}$ denotes the set
$\{\nu\pm\nu'\,|\,\nu\in\CC^n_{\bf k},\nu'\in\CC^n_{\bf k'}\}$.
Note that $\Omega_n(K)\subset\Omega_n(K')$ whenever $K>K'$.
Furthermore, one introduces for $\omega\in\real^d$ the subsets of
$\integer^d$
$$
\QQ^+_\omega=\{q\in\integer^d\,|\,\omega\cdot q>0\}\,,\quad
\QQ^-_\omega=\{q\in\integer^d\,|\,\omega\cdot q<0\}.
\equation(defqpqm)
$$

\claim Proposition(itbounds)
There exist positive constants $r$ and $\lambda_0$
small enough such that
the following is true
for $|\lambda|<\lambda_0$, $n\geq1$, and $|\Im\beta|<\alpha_n$, where
$\alpha_1=\alpha$ and, for $n\geq2$,
$$
\alpha_n=(1-n^{-2})\alpha_{n-1}.
\equation(defalphan)
$$
There exists $K_\lambda>0$ satisfying $K_\lambda\rightarrow0$ as
$\lambda\rightarrow0$
such that one has for
$\omega\in\Omega_n(K_\lambda)$ arbitrary but fixed,
\item{\rm(a)} Equation \equ(eqrnb) has a solution
$R_{n\beta}$ in $H^\infty(B_n,h^{n-1}_{s})$
analytic in $|\lambda|<\lambda_0$ and $(\Phi,J)\in\sO_\sB$.
\item{\rm(b)} Defining $w_{n\beta}$ according to
\equ(defwnb), one has $w_{n\beta}\in\AA_n$ and, writing
$w_{n\beta}(z)\equiv w_n(z)=w_n(0)+Dw_n(0)z
+\delta_2w_n(z)$,
$$\eqalignno{
&||\hat P_nw_n(0)||_{s'}\leq\varepsilon r^{2n},&\equat(fstbnd)\cr
&|||\hat P_n\delta_2w_n|||_{\AA_n}\leq\varepsilon r^{2n},
&\equat(sndbnd)\cr
}$$
\item{}where $\varepsilon\rightarrow0$ as $\lambda\rightarrow0$.
\item{\rm(c)} There exists $A_n\in\LL(h_s,h_{s'})$
such that $\tilde w_n\equiv w_n-A_n$ obeys for all $z\in B_n$,
$$
||\hat P_nD\tilde w_n(z)||^{(n)}_{s,s'}\leq\varepsilon\eta^{n}.
\equation(thdbnd)
$$
Furthermore,
$$
||A_n||_{s,s'}\leq3\varepsilon\eta^{n-1},
\equation(deltan)
$$
$A_n(q,q')=0$ if $q\not=q'$ and
$$
A_n(q,q)=a_n{\rm I}_{\QQ^+_\omega}(q)+
\overline{a_n}\,\,{\rm I}_{\QQ^-_\omega}(q),
\equation(defGn)
$$
where
$a_n\in\LL(\hat\RR_s^\infty,\hat\RR_{s'}^\infty)$
is hermitian, {i.e.}, $a_n=\overline{a_n}^{\rm T}$,
and satisfies for all ${\bf k}\in\II^n$,
$$
a_n \JJ^n_{\bf k}=\JJ^n_{\bf k}\,.
\equation(subinv)\
$$
\item{\rm(d)} The matrix
$\tilde\mu_{n+1}^2\equiv\mu^2+\sum_{m=0}^{n}a_m$
is positive definite and the spectrum of
$\tilde\mu_{n+1}$ can be uniquely decomposed
into a maximal family of pairwise
disjoint clusters $\CC^{n+1}_{k,i}$, $k\geq1$, $i=1,\dots,M^{n+1}_k$,
with $M^{n+1}_k\geq M^{n}_k$, satisfying for all $k\geq1$
the gap condition
$$
d(\CC^{n+1}_{k,i},\CC^{n+1}_{k,j})>\eta^{n+1}
\quad{\rm if}\quad i\not=j,
\equation(gapn2)
$$
and
$$
\nu=\mu_k+\OO(\varepsilon k^{-\xi})
\quad{\rm for\ all}\quad
\nu\in\CC^{n+1}_{k,i}\,,\ i=1,\dots,M_k^{n+1}.
\equation(asymp)
$$
Furthermore,
the sets $S^{n+1}_{k,i}$ defined according to
\equ(defsin) are pairwise disjoint, and \equ(convclust), \equ(PkiPlj)
and \equ(proppn) hold with $n$
replaced by $n+1$.
\item{}

Let us briefly comment on \clm(itbounds), whose proof will be
carried out in Section~6.
First, we note that point (d) ensures, in particular,
that the new set of clusters
$\CC^{n+1}_{k,i}$ enjoy the properties required for proceeding to the
next step of the induction, cf. the discussion at the end of
Section~3. The asymptotic behavior \equ(asymp) concerns the
measure estimate of the set $\Omega^*$ of admissible frequencies
in \clm(mainthm). Such an asymptotic behavior is required in order
to obtain a
set of large measure because one imposes Diophantine conditions with
respect to differences of the normal frequencies.
We will show in Section~7 that
\equ(asymp) implies the

\claim Proposition(measureest)
For $\nu=\nu(d,\xi)$ sufficiently large, the set
$$
\Omega^*(K)\equiv\bigcap_{n\geq1}\Omega_n(K)
\equation(omegastar)
$$
satisfies
for all
bounded $\Omega\subset\real^d$,
${\rm meas}(\Omega\setminus\Omega^*(K))\rightarrow0$
as $K\rightarrow0$.

Note that $\omega\in\Omega^*$ assume a Diophantine condition with
respect to zero. Therefore, one has for such $\omega$,
$\integer^d\setminus\{0\}=\QQ^+_\omega\cup\QQ^-_\omega$.
Next, we turn to bound \equ(thdbnd), the most delicate estimate
to establish.
To treat the off-diagonal part $Dw_n(q,q')$, $q\not=q'$, we will
rely on the fact that the exponential decay of the kernel
$Dw_0(q,q')$ in the size of $|q-q'|$ is
preserved due
to the introduction of the parameter $\beta$.
We note that imposing Diophantine conditions on $\omega$ with
respect to the differences $\CC^n_{\bf k}\pm\CC^n_{\bf k'}$
ensures that
$|q-q'|$ is of order $\OO(\eta^{-n/\nu})$ for $q\not=q'\in S_n$.
To treat the diagonal part, we will use that
$Dw_n(q,q)$ depends on $q$ through $\omega\cdot q$
only, and is, in some sense, continuous in this variable. More
precisely, defining
$t_p:\LL(h_s,h_{s'})\rightarrow\LL(h_s,h_{s'})$, $p\in\integer^d$,
by
$$
(t_p L)(q,q')=L(q+p,q'+p),
\equation(deftp)
$$
and setting
$$
\Delta_p\equiv t_p-\identity,
\equation(defDeltap)
$$
we will show that $\Delta_p Dw_n$ is of order
$\OO(\varepsilon|\omega\cdot p|)$ on the diagonal.
Therefore, since $p=q-q'$ satisfies
$|\omega\cdot p|\leq\eta^n$
for $q,q'\in S^n_{\bf k}$ such that
$\sign(\omega\cdot q)=\sign(\omega\cdot q')$,
one has for $q\in S^n_{\bf k}$,
$$
\hat P_nDw_n(q,q)\hat P_n=\hat a_{\bf k}+\OO(\varepsilon\eta^{n}),
$$
where $\hat a_{\bf k}:\JJ^n_{\bf k}\rightarrow\JJ^n_{\bf k}$
dependents only on the sign of $\omega\cdot q$.
The continuity of $Dw_n(q,q)$ ultimately follows from the fact that
$\Gamma_n(q)$
is continuous in $\omega\cdot q$, as stated in the following lemma,
whose proof can be found in the Appendix.
\claim Lemma(intermediary)
Let $\sigma\in\real$ and $p\in\integer^d$. Then
the operator $\Gamma_n=\sK_n^{-1}Q_nP_{n-1}$ obeys
$$\eqalignno{
||\Gamma_n||_{\sigma,\sigma+\gamma}&\leq C\eta^{-n},
&\equat(bndinvan)\cr
||\Delta_p\Gamma_n||_{\sigma,\sigma+\gamma}&\leq
C\eta^{-2n}|\omega\cdot p|.
&\equat(DeltapG)\cr
}$$

Finally, the perturbation $a_n$ being hermitian will essentially
follow from the reality of the original equation \equ(sys2).
More precisely,
the derivative $Dw_n$ satisfies
$$\eqalignno{
Dw_n^{ij}(q,q')&=\overline{Dw_n^{ij}(-q,-q')},&\equat(symm1)\cr
Dw_n^{ij}(q,q')&=Dw_n^{ji}(-q',-q).&\equat(symm2)\cr
}$$
Thus, the diagonal element
$Dw_n(q,q):\hat\RR^\infty\rightarrow\hat\RR^\infty$ is given by an
hermitian matrix for all $q$, and $a_n$ hermitian will follow since,
as was mentioned above, $a_n$ will be chosen in such a way that its
action on each $\JJ^n_{\bf k}$ is the constant approximation of
$Dw_n(q,q)$ for $q\in S^n_{\bf k}$.
Note that due to \equ(symm1), one expects $Dw_n(-q,-q)$ to be
approximated by $\overline{a_n}$, which explains the decomposition in
formula \equ(defGn).
Identities \equ(symm1) and \equ(symm2) are easily checked to hold for
$n=0$. Indeed, the perturbation $U$
in the Hamiltonian \equ(hamiltonian) being real analytic ensures
\equ(symm1), whereas \equ(symm2) follows from the fact that $Dw_0$ is
the symmetric second derivative of the functional
$Z\mapsto\lambda\int U(\varphi+\Phi(\varphi),J(\varphi),
Z(\varphi))d\varphi$,
cf. \equ(wrho). Using the recursive relations \equ(defwn) and
\equ(defwtilden), one obtains \equ(symm1) and \equ(symm2) for $n\geq1$
by iteration.

\claimrm Remark(rem1) The choice of constants is as follows.
We first fix $\eta$ small enough according, essentially, to the
constants entering the asymptotics of the frequencies $\mu_k$ in (H1),
cf. Section~6.4. Given $\eta$, $\varepsilon$ and $r$ are chosen
small enough, and $\lambda_0$ is chosen in turn according to
$\varepsilon$. The latter choice plays a role only in ensuring that
the inductive hypothesis of \clm(itbounds) are satisfied for $n=0$,
cf. the introduction in Section~6. Finally, $K_\lambda$ is chosen
large enough in order for the estimate
$$
Ce^{-Cn^{-2}K_\lambda^{1/\nu}\eta^{-n/\nu}}\leq r^{2n},
\equation(expest)
$$
to hold for all $n\geq1$. This will be needed in order to iterate the
bound \equ(fstbnd) in Section~6.2. Note that due to the double
exponential, the dependence of $K_\lambda$ on $\eta$ and $r$ is given
by the behavior at small $n$ of the expressions entering
\equ(expest). That $K_\lambda$ can be taken smaller as
$\lambda$ goes to zero will follow from the fact that $r$ and
$\varepsilon$, and thus ultimately $\eta$, can be taken smaller.
Finally, we denote by $C$ a generic constant, independent on $n$, $r$,
and $\varepsilon$,
which may vary from place to place.

\section Proof of \clm(itbounds)

We proceed by induction and assume that \clm(itbounds) holds up to
$n-1\geq1$. Regarding the inductive hypothesis in the
case $n=1$, we simply choose $A_0\equiv0$, so that the bounds for
$w_0$ in points (b) and (c) of \clm(itbounds) are a simple
consequence of \equ(nw0beta). Furthermore, $\tilde\mu_1=\mu$ and point
(d) follows immediately from (H1). We note that in Section~6.1 below,
point (a) is established for $n=1$ by taking $\varepsilon$,
namely $\lambda$, small
enough. At some point in the induction, however, one is forced to
consider nontrivial $A_n$ in order for the inductive
bounds to hold uniformly in $n$ for a given $\lambda$.

In the sequel, we adopt the convention, for
$B$ a ball of radius $r$ centered at the origin, to denote by $\gamma
B$ the ball of radius $\gamma r$ centered at the origin.

\subsection Existence of the Functional $R_{n\beta}$

With the notations $R=R_{n\beta}$,
$\Gamma=\Gamma_n$ and $\tilde w=\tilde w_{(n-1)\beta}$,
equation \equ(eqrnb) reads
$$
R(z)=\Gamma\tilde w(z+R(z)).
\equation(equrn2)
$$
To prove existence in $H^\infty(B_{n},h^{n-1}_{s})$ of a
solution $R$ to equation \equ(equrn2),
one starts,
using the identities $\tilde w(0)=w(0)$ and
$\delta_2\tilde w=\delta_2w$,
by decomposing $\tilde w$ as
$$
\tilde w(z)=w(0)+D\tilde w(0)z
+\delta_2w(z),
\equation(decomposition)
$$
to obtain from \equ(equrn2),
$$
R(z)=\Gamma w(0)
+\Gamma D\tilde w(0)(z+R(z))
+\Gamma\delta_2w(z+R(z)).
\equation(equR)
$$
Defining
$$
H=\bigl(1-\Gamma D\tilde w(0)\bigr)^{-1},
\equation(defh)
$$
and using the identity $1+H\Gamma D\tilde w(0)=H$,
one rewrites \equ(equR) as
$$
R(z)=H\Gamma w(0)
+H\Gamma D\tilde w(0)z
+u(z),
\equation(equwtilde2)
$$
where
$$
u(z)=H\Gamma\delta_2 w(\tilde z)
\equiv\GG(u)(z),
\equation(equu)
$$
and
$$
\tilde z\equiv z+R(z)=H\bigl(z+\Gamma w(0)\bigr)+u(z).
\equation(ytilde)
$$
Since $\Gamma=\Gamma\hat P_{n-1}=\hat P_{n-1}\Gamma$,
\equ(bndinvan) (with $\sigma=s+\xi-\gamma$)
and the recursive bound \equ(thdbnd)
(with $n$ replaced by $n-1$) imply
$$
||\Gamma D\tilde w(0)||^{(n-1)}_{s}\leq
||\Gamma D\tilde w(0)||^{(n-1)}_{s,s+\xi}\leq C\varepsilon\eta^{-1}.
\equation(int1)
$$
Hence,
$$
||H||^{(n-1)}_s\leq2,
\equation(boundonh)
$$
for $\varepsilon=\varepsilon(\eta)$ small enough.
Since
$B_{n}\subset B_{n-1}$, $\tilde w(0)=w(0)$, and since bounds
\equ(fstbnd) (with $n$ replaced by $n-1$),
\equ(bndinvan)
and \equ(int1) hold,
the existence of $R$ in $H^\infty(B_{n},h^{n-1}_s)$ follows from
the existence of $u$ in $H^\infty(B_{n},h^{n-1}_s)$.
For reasons that will become clear in the next section, we actually
show that \equ(equu) has a solution $u$ in the ball
$$
\BB=\Bigl\{u\in H^\infty(\oneeighth B_{n-1},h^{n-1}_s)\ |\
|||u|||\leq\sqrt{\varepsilon}\eta^{-n}r^{2(n-1)}\Bigr\}.
\equation(ballBB)
$$
This result is stronger, since
$B_{n}\subset\oneeighth B_{n-1}$ for $r$ small enough.
Let us first check that $\GG$ maps $\BB$ into
itself. From \equ(boundonh) and the recursive bound \equ(fstbnd),
it follows that for all $z\in\oneeighth B_{n-1}$ and $u\in\BB$,
$\tilde z\in h^{n-1}_s$ with
$$
||\tilde z||_s\leq
2(\oneeighth r^{n}+C\varepsilon\eta^{-n}r^{2(n-1)})
+\sqrt{\varepsilon}\eta^{-n}r^{2(n-1)}
\leq\onehalf r^{n},
$$
for $\varepsilon=\varepsilon(r,\eta)$
and $r=r(\eta)$ small enough. Hence,
$$
\tilde z\in\onehalf B_{n-1}\subset B_{n-1}
\quad{\rm for\ all\ }z\in\oneeighth B_{n-1},
\equation(tildeyinbn1)
$$
and one uses the bound \equ(sndbnd)
to conclude that
for all $u\in\BB$,
$$
|||\GG(u)|||\leq2C\eta^{-n}\varepsilon r^{2(n-1)}
\leq\sqrt{\varepsilon}\eta^{-n}r^{2(n-1)},
$$
for $\varepsilon$ small enough.
To show that
$\GG$ is a contraction in $\BB$, we apply the estimate \equ(Cauchy2)
to the functions $\tilde z_i$ given by \equ(ytilde) in terms of
$u_i\in\BB$, $i=1,2$. Noting that
$|||\tilde z_i|||\leq\onehalf r^{n}$,
which follows from \equ(tildeyinbn1),
and using in addition \equ(sndbnd), one obtains,
$$\eqalign{
|||\GG(u_1)-\GG(u_2)|||&\leq2C\eta^{-n}
\sup_{z\in\oneeighth B_{n-1}}||\hat P_{n-1}\delta_2w(\tilde z_1)-
\hat P_{n-1}\delta_2w(\tilde z_2)||_{s'}\cr
&\leq4C\eta^{-n}r^{-n}|||\hat P_{n-1}\delta_2w|||_{\AA_{n-1}}
\sup_{z\in\oneeighth B_{n-1}}||\tilde z_1-\tilde z_2||_s\cr
&\leq4C\varepsilon\eta^{-n}r^{-n}r^{2(n-1)}
\sup_{z\in\oneeighth B_{n-1}}||u_1(z)-u_2(z)||_s\cr
&\leq\onehalf|||u_1-u_2|||,\cr
}$$
for $r=r(\eta)$ and $\varepsilon=\varepsilon(r,\eta)$ small enough.

Before turning to part (b) of \clm(itbounds), we make some remarks
that shall be useful later.
First note that
\equ(tildeyinbn1) means
$$
z+R_n(z)\in\onehalf B_{n-1}
\quad{\rm for\ all}\quad z\in\oneeighth B_{n-1}.
\equation(tildeyinbn)
$$
Therefore, with
$$
\tilde R_m(z)\equiv z+R_m(z),
\equation(deftildeRn)
$$
and
$$
F_n^m(z)\equiv\tilde R_m\circ\tilde R_{m+1}\circ
\dots\circ\tilde R_n(z),
\equation(deffnm)
$$
it follows recursively that for $m=1,\dots,n$,
$$
F_n^m(z)\in\onehalf B_{m-1}
\quad{\rm for\ all}\quad z\in B_{n}.
\equation(fnmin)
$$
Furthermore, since $F_n^1=F_n$, where $F_n$ is defined in \equ(deffn),
one has $F_n\in\AA_n$, together with the uniform bound
$$
|||F_n|||_{\AA_n}\leq|||\tilde R_1|||_{\AA_1}\leq\varepsilon.
\equation(unifFn)
$$

\subsection Bounds on the Functional $w_n$

According to \equ(defwnb), one defines
$$
w_{n\beta}(z)=\tilde w_{(n-1)\beta}(z+R_{n\beta}(z)).
$$
Since $R_{n\beta}\in H^\infty(B_{n},h^{n-1}_s)$,
it follows from \equ(tildeyinbn) and the
inductive bounds that for all $\beta$ with $|\Im\beta|<\alpha_{n-1}$,
$w_{n\beta}$ is well defined as a map from $B_n$ to
$h_{s'}$, with $w_{n\beta}\in\AA_n$.
In the sequel, we adopt the simplified notation $R=R_{n\beta}$,
$w=w_{(n-1)\beta}$ and $w'=w_{n\beta}$.
We proceed with proving \equ(fstbnd).
Using the decomposition \equ(decomposition) at $z=0$, one may write
$$
w'(0)=w(0)+D\tilde w(0)R(0)+\delta_2w(R(0)).
$$
Since \equ(tildeyinbn) implies that
$R(0)\in\onehalf B_{n-1}$,
one obtains using the bounds \equ(fstbnd), \equ(sndbnd) and
\equ(thdbnd),
$$\eqalignno{
||\hat P_nw'(0)||_{s'}&\leq\varepsilon r^{2(n-1)}
+\onehalf\varepsilon\eta^{n-1}r^{n}+
\varepsilon r^{2(n-1)}&\cr
&\leq3\varepsilon.&\equat(pbjunk1)\cr
}$$
This leads to
$$
|\PP^n_{\bf k}w'(0)(q)|_{s'}\leq3\varepsilon,
\equation(pbjunk2)
$$
for all
${\bf k}\in\II^n$ and $q\in S^n_{\bf k}$.
The latter is valid for all $\beta$ with
$|\Im\beta|<\alpha_{n-1}$.
Let now $\beta'$ with $|\Im\beta'|<\alpha_{n}$.
Then, shifting $\beta'$ to
$\beta=\beta'-i(\alpha_{n-1}-\alpha_{n})q/|q|$ and using the recursive
relation \equ(defalphan) for $\alpha_n$, one obtains
$$
w_{\beta'}'(0)(q)=e^{i(\beta'-\beta)\cdot q}w_{\beta}'(0)(q)
=e^{-n^{-2}\alpha_{n-1}|q|}w_{\beta}'(0)(q).
\equation(betabetap)
$$
Since for such $\beta'$ one has $|\Im\beta|<\alpha_{n-1}$,
it follows from
\equ(pbjunk2) and \equ(betabetap) that
$$\eqalignno{
||\hat P_nw'(0)||_{s'}&\leq3\varepsilon\sum_{{\bf k}\in\II^n}
\sum_{q\in S^n_{\bf k}}e^{-n^{-2}\alpha_{n-1}|q|}\ .
&\equat(boundonwp)\cr
}$$ From the Diophantine conditions
satisfied by $\omega\in\Omega_n(K)$, one infers for
$q\in S^n_{\bf k}$
that $|q|>\min(K\eta^{-n/\nu},(4K)^{1/\nu}\eta^{-n/\nu})$,
cf. \equ(defsin) and \equ(omegank). Therefore, Bound \equ(fstbnd)
finally follows by choosing $K$ appropriately, cf. \equ(expest).

We now iterate bound \equ(sndbnd).
Using again the decomposition \equ(decomposition),
one has
$$\eqalign{
\delta_2 w'(z)&=D\tilde w(0)\delta_2R(z)
+\delta_2\delta_2w(z+R(z)).\cr
}$$
The first term on the right hand side is estimated by using
$\delta_2R(z)=\delta_2u(z)$ together with \equ(Cauchy3) applied to
$u\in\BB$ with $\gamma=8r$, since $B_{n}\subset\oneeighth B_{n-1}$,
to obtain
$$\eqalign{
|||\hat P_nD\tilde w(0)\delta_2R|||_{\AA_{n}}
&\leq||\hat P_{n-1}D\tilde w(0)||^{n-1}_{s,s'}
\sup_{z\in B_{n}}||\delta_2 u(z)||_s\cr
&\leq\varepsilon\eta^n{(8r)^2\over1-(8r)^2}|||u|||\cr
&\leq\varepsilon r^{2n}{\sqrt{\varepsilon}\,8^2\over1-(8r)^2}\cr
&\leq\onehalf\varepsilon r^{2n},
}$$
for $\varepsilon$ small enough. In a similar way, one estimates, using
\equ(tildeyinbn), that
$$
\sup_{z\in B_n}||\hat P_n\delta_2\delta_2w(z+R(z))||_{s'}
\leq\onehalf\varepsilon r^{2n},
$$
which finally leads to \equ(sndbnd).

\subsection Bounds on the Derivative

In this section, we prove the estimates stated in part (c)
of \clm(itbounds).
The main difficulty consists in controlling the diagonal part of the
kernel of the derivative $Dw_n$ evaluated at zero, namely
$Dw_n(0)(q,q)$, $q\in\integer^d$.
To address this problem, as mentioned in the end of Section~5,
we will use the fact that $Dw_n(0)(q,q)$ depends
on $q$ through $\omega\cdot q$ only, and satisfies some continuity
property when viewed as a function of $\omega\cdot q$.

We start by deriving an a priori bound on the norm
of $Dw_n$. From \equ(equrn), one infers that
$$
DR_n(z)=H_n(\tilde z)\Gamma_n D\tilde w_{n-1}(\tilde z),
\equation(DRn)
$$
where
$$\eqalignno{
H_n(\tilde z)&=\bigl(1-\Gamma_n D\tilde w_{n-1}(\tilde z)\bigr)^{-1},
&\equat(defH)\cr
\tilde z&=z+R_n(z).&\equat(defytilde)\cr
}$$
Since by definition, cf. \equ(defwn), one has
$$
Dw_n(z)=D\tilde w_{n-1}(\tilde z)\bigl(1+DR_n(z)\bigr),
$$
\equ(DRn) and the identity
$H_n(\tilde z)=1+H_n(\tilde z)\Gamma_n D\tilde w_{n-1}(\tilde z)$,
imply the recursive relation
$$
Dw_n(z)=D\tilde w_{n-1}(\tilde z)H_n(\tilde z).
\equation(recrelDw)
$$
As in the previous section, it follows from \equ(bndinvan),
\equ(tildeyinbn), and the
inductive bounds, that
$||H_n(\tilde z)||^{(n-1)}_s\leq2$ for all $\tilde z\in B_{n-1}$.
Therefore, one obtains
for all $z\in\oneeighth B_{n-1}$,
using again the inductive bound \equ(thdbnd),
$$
||\hat P_nDw_n(z)||^{(n)}_{s,s'}\leq
||\hat P_{n-1}Dw_n(z)||^{(n-1)}_{s,s'}\leq2\varepsilon\eta^{n-1}.
\equation(aprioriDw)
$$

In order to iterate bounds \equ(thdbnd),
we decompose $Dw_n(z)$ as follows
$$
Dw_n(z)=\sigma_n+\tau_n+\delta_1Dw_n(z),
\equation(decompDwn)
$$
where $\sigma_n+\tau_n=Dw_n(0)$ and
$\sigma_n(q,q')=Dw_n(0)(q,q')\delta_{qq'}$. Let us consider first the
last two terms on the right hand side of \equ(decompDwn). One has the

\claim Lemma(delta1wn) Let $r$ and $\varepsilon$ be the
positive constants of \clm(itbounds). Then, one has for
all $n\geq0$ and all $z\in B_n$,
$$\eqalignno{
||\hat P_n\delta_1Dw_n(z)||^{(n)}_{s,s'}&
\leq\onehalf\varepsilon r^{{n\over2}},
&\equat(boundd1Dwn)\cr
||\hat P_n\tau_n||^{(n)}_{s,s'}&\leq\varepsilon r^n.
&\equat(bndtaufinal)\cr
}$$

\proof Proceeding by induction, we suppose that \clm(itbounds) and
\clm(delta1wn) are true up to some $n-1$, $n\geq1$.
We start with \equ(boundd1Dwn) and compute from
$\delta_1Dw_n(z)=Dw_n(z)-Dw_n(0)$ and the recursive relation
\equ(recrelDw) that
$$
\delta_1Dw_n(z)=
\tilde H_n(\tilde z_0)
\bigl(D\tilde w_{n-1}(\tilde z)-D\tilde w_{n-1}(\tilde z_0)\bigr)
H_n(\tilde z),
$$
where $\tilde z_0=R_n(0)$ and
$\tilde H_n(\tilde z_0)=1+Dw_{n-1}(\tilde z_0)
H_n(\tilde z_0)\Gamma_n$.
As previously, the inductive bound \equ(thdbnd)
implies $||\hat P_n\tilde H_n(\tilde z_0)||^{(n-1)}_{s'}\leq2$.
Using \equ(tildeyinbn) and
$\hat P_n\tilde H_n=\hat P_n\tilde H_n\hat P_{n-1}$,
one infers from the identity
$D\tilde w_{n-1}(\tilde z)-D\tilde w_{n-1}(\tilde z_0)=
\delta_1D\tilde w_{n-1}(\tilde z)-\delta_1D\tilde w_{n-1}(\tilde z_0)$
that for all $z\in\oneeighth B_{n-1}$,
$$
||\hat P_n\delta_1Dw_n(z)||^{(n-1,n)}_{s,s'}
\leq C\sup_{z'\in\onehalf B_{n-1}}||\hat P_{n-1}\delta_1
D\tilde w_{n-1}(z')||^{(n-1)}_{s,s'}.
$$
Since $\delta_1D\tilde w_{n-1}=\delta_1Dw_{n-1}$,
the recursive bound \equ(boundd1Dwn) leads to
$$
||\hat P_n\delta_1Dw_n(z)||^{(n-1,n)}_{s,s'}
\leq C\varepsilon r^{{n-1\over2}},
$$
for all $z\in\oneeighth B_{n-1}$.
Finally, iterating bound \equ(boundd1Dwn) is completed by restricting
$z$ to $B_n\subset\oneeighth B_{n-1}$ and using \equ(Cauchy3) with
$\gamma=8r$.

Next, we turn to \equ(bndtaufinal), the estimate for
the off-diagonal part of $Dw_n(0)$.
The norm of $\tau_n$ reads
$$
||\hat P_n\tau_n||^{(n)}_{s,s'}=
\sup_{{\bf k'}\in\II^n}\sup_{q'\in S^n_{\bf k'}}
\sum_{{\bf k}\in\II^n}\sum_{q\in S^n_{\bf k}}
|\PP^n_{\bf k}\tau_n(q,q')\PP^n_{\bf k'}|_{s,s'}\,,
$$
and one infers from \equ(boundd1Dwn) and
the a priori bound \equ(aprioriDw) that
$$
|\PP^n_{\bf k}\tau_n(q,q')\PP^n_{\bf k'}|_{s,s'}
\leq2\varepsilon\eta^{n-1}+\onehalf\varepsilon r^{n\over2}
\leq3\varepsilon.
\equation(appitnpj)
$$
The latter is valid for all $\beta$ with
$|\Im\beta|<\alpha_{n-1}$.
Let now $\beta'$ with $|\Im\beta'|<\alpha_{n}$.
Then, shifting $\beta'$ to
$\beta=\beta'-i(\alpha_{n-1}-\alpha_{n})(q-q')/|q-q'|$, one obtains
$$
\tau_{n\beta'}(q,q')=e^{i(\beta'-\beta)\cdot(q-q')}\tau_{n\beta}(q,q')
=e^{-n^{-2}\alpha_{n-1}|q-q'|}\tau_{n\beta}(q,q').
\equation(bbprime)
$$
Hence, since $|\Im\beta|<\alpha_{n-1}$ for such $\beta'$,
\equ(appitnpj) and \equ(bbprime) lead to
$$
||\hat P_n\tau_{n}||^n_{s,s'}\leq3\varepsilon
\sup_{{\bf k'}\in\II^n}\sup_{q'\in S^n_{\bf k'}}
\sum_{{\bf k}\in\II^n}
\sum_{\scriptstyle q\in S^n_{\bf k}\atop\scriptstyle q\not=q'}
e^{-n^{-2}\alpha_{n-1}|q-q'|}\ .
\equation(bndtau1)
$$
We now show that every term in the previous sum yields a
super-exponentially small
factor. Let $q\in S^n_{\bf k}$ and $q'\in S^n_{\bf k'}$ for some
${\bf k}\in\II^n$, ${\bf k'}\in\II^n$.
Then, one estimates using \equ(defsin) and \equ(bndabsIin)
that if $\sign(\omega\cdot q)=\sign(\omega\cdot q')$,
$$
d\bigl(|\omega\cdot(q-q')|,\CC^n_{\bf k}+\CC^n_{\bf k'}\bigr)\leq\
\onehalf\eta^n+|I^n_{\bf k}|+|I^n_{\bf k'}|
\leq3\bar d\eta^n,
$$
and that otherwise
$$
d\bigl(|\omega\cdot(q-q')|,|\CC^n_{\bf k}-\CC^n_{\bf k'}|\bigr)\leq\
\onehalf\eta^n+|I^n_{\bf k}|+|I^n_{\bf k'}|
\leq3\bar d\eta^n.
$$
Therefore, since $q\not=q'$, it follows from \equ(omegank)
and  $\omega\in\Omega_n(K)$ that
$$
|q-q'|\geq\min\Bigl(\Bigl({K\over3\bar d}\Bigr)^{1/\nu},K\Bigr)
\eta^{-n/\nu}.
$$
Hence, the contribution of each term in \equ(bndtau1) is
super-exponentially small, and \equ(bndtaufinal) follows
for some $r\ll\eta<1$.
\qed

Finally, we turn to $\sigma_n$, the diagonal part of $Dw_n(0)$ in the
decomposition \equ(decompDwn).
We first state a result about
the continuity properties of the kernel $\sigma_n(q,q)$,
namely that
$\Delta_p\sigma_n=t_p\sigma_n-\sigma_n$
is of order $|\omega\cdot p|$.
More precisely, one has the
\claim Proposition(continuity)
Suppose that \clm(itbounds) is valid up to $n-1$ for some
$n\geq1$. Then, the diagonal part $\sigma_n(z)$ of $Dw_n(z)$ satisfies
for all $z\in B_n$ and all $p$ such that
$|\omega\cdot p|<{1\over16}\eta^{n-1}$,
$$\eqalignno{
||\hat P_n\Delta_p\sigma_n(z)||^{n}_{s,s'}
&\leq\varepsilon^{{3\over2}}|\omega\cdot p|.
&\equat(Dpsigman)\cr
}$$

Delaying the proof of the above proposition to the end of this
section, we now
construct a diagonal operator $A_n\in\LL(h_s,h_{s'})$
such that
$\tilde\sigma_n\equiv\sigma_n-A_n$ obeys
$$
||\hat P_n\tilde\sigma_n||^n_{s,s'}=
\sup_{{\bf k}\in\II^n}\sup_{q\in S^n_{\bf k}}
|\PP^n_{\bf k}\tilde\sigma_n(q,q)\PP^n_{\bf k}|_{s,s'}
\leq\onehalf\varepsilon\eta^{n}.
\equation(bndtildesigma)
$$
The equality above follows from the sets $S^n_{\bf k}$
being pairwise disjoint.
This will conclude the proof of iterating \equ(thdbnd), since
\equ(boundd1Dwn), \equ(bndtaufinal) and \equ(bndtildesigma) imply
that the derivative of $\tilde w_n\equiv w_n-A_n$ satisfies the
required bound for $r=r(\eta)$ small enough.
In order to obtain bound \equ(bndtildesigma) by using the continuity
property \equ(Dpsigman), we would like to construct $A_n$ as an
approximation of $\sigma_n(q,q)$ for $\omega\cdot q$ close to the
normal frequencies in $\CC^n_{\bf k}$, ${\bf k}\in\II^n$.
To this end, we set $\bar\mu_{\bf k}$ to be the center
of the interval $I^n_{\bf k}$ and, using that
$\{\omega\cdot q\,|\,q\in\integer^d\}$ is dense in $\real$,
we choose a sequence
$\{q_{l,{\bf k}}\}_{l\geq1}\subset S^n_{\bf k}$
such that $\omega\cdot q_{l,{\bf k}}>0$ for all $l\geq1$ and
$$
\lim_{l\rightarrow\infty}\omega\cdot q_{l,{\bf k}}=\bar\mu_{\bf k}.
$$
Next, one defines the matrix
$\hat a_{n,{\bf k}}\in\LL(\JJ^n_{\bf k})$ by
$$
\hat a_{n,{\bf k}}\equiv\lim_{l\rightarrow\infty}
\PP^n_{\bf k}\sigma_n(q_{l,{\bf k}},q_{l,{\bf k}})\PP^n_{\bf k}.
\equation(deltain)
$$
Due to \equ(Dpsigman), the limit in \equ(deltain) exists and
does not depend on the particular choice of the sequence
$\{q_{l,{\bf k}}\}_{l\geq1}$.
Finally, setting
$$
a_n\equiv
\bigoplus_{{\bf k}\in\II^n}
\hat a_{n,{\bf k}},
\equation(defsigman0)
$$
we define the operator $A_n:h\rightarrow h$
as given by the diagonal kernel
$$
A_n(q,q)=a_n{\rm I}_{\QQ^+_\omega}(q)+
\overline{a_n}\,\,{\rm I}_{\QQ^-_\omega}(q)
\equation(defsigman)
$$
for all $q\in\integer^d$.
We note that by construction, \equ(subinv) is clearly satisfied.
Furthermore, it follows from \equ(symm1) and \equ(symm2)
that $a_n$ is indeed hermitian.
Let us check that definition \equ(defsigman) leads to the required
bound \equ(bndtildesigma). By construction, one has for all
${\bf k}\in\II^n$,
$$
\lim_{l\rightarrow\infty}\PP^n_{\bf k}
\tilde\sigma_n(q_{l,{\bf k}},q_{l,{\bf k}})\PP^n_{\bf k}=0.
\equation(limtilde)
$$
On the other hand, since $\Delta_pA_n=0$, bound \equ(Dpsigman) is
also satisfied by $\tilde\sigma_n$, which by definition of the norm
implies that
$$
|\PP^n_{\bf k}\Delta_p\tilde\sigma_n(q,q)\PP^n_{\bf k}|_{s,s'}\leq
\varepsilon^{{3\over2}}|\omega\cdot p|,
\equation(contts)
$$
for all $q\in S^n_{\bf k}$, ${\bf k}\in\II^n$,
and $p\in\integer^d$ with $|\omega\cdot p|<{1\over16}\eta^{n-1}$.
The definition of
$S^n_{\bf k}$ together with \equ(bndabsIin) implies that
$|\omega\cdot(q-q')|\leq2\bar d\eta^n\leq{1\over16}\eta^{n-1}$
for all $q,q'\in S^n_{\bf k}$ with
$\sign{(\omega\cdot q)}=\sign{(\omega\cdot q')}$ and $\eta$
small enough. Therefore, using
$$
\tilde\sigma_n(q,q)=
\tilde\sigma_n(q',q')+
\Delta_{q-q'}\tilde\sigma_n(q',q'),
$$
one infers from \equ(contts)
that for all $q_{l,{\bf k}}$ and $q\in S^n_{\bf k}$ with
$\omega\cdot q>0$,
$$
|\PP^n_{\bf k}\tilde\sigma_n(q,q)\PP^n_{\bf k}|_{s,s'}\leq
|\PP^n_{\bf k}\tilde\sigma_n(q_{l,{\bf k}},q_{l,{\bf k}})
\PP^n_{\bf k}|_{s,s'}+
\varepsilon^{{3\over2}}|\omega\cdot(q-q_{l,{\bf k}})|,
\equation(tsigtmp)
$$
which, with \equ(limtilde), leads to
$$
|\PP^n_{\bf k}\tilde\sigma_n(q,q)\PP^n_{\bf k}|_{s,s'}\leq
2\bar d\varepsilon^{{3\over2}}\eta^n.
\equation(fbnd)
$$
For $q\in S^n_{\bf k}$ with $\omega\cdot q<0$,
we note that \equ(tsigtmp) is also valid if one replaces
$q_{l,{\bf k}}$ by $-q_{l,{\bf k}}$, and, due to \equ(symm1),
that the same
is true of \equ(limtilde). Therefore, \equ(fbnd) holds for all
$q\in S^n_{\bf k}$, ${\bf k}\in\II^n$, and
bound \equ(bndtildesigma) follows by taking
$\varepsilon$ small enough.
Finally, we check that $A_n$ obeys \equ(deltan).
The a priori bound \equ(aprioriDw) together with \equ(bndtildesigma)
imply that $||\hat P_nA_n||^{(n)}_{s,s'}\leq3\varepsilon\eta^{n-1}$,
which, with \equ(subinv) and definition \equ(defsigman),
leads to \equ(deltan).

To complete the proof of part (c) of \clm(itbounds),
we are left with the

\noindent{\bf Proof of \clm(continuity).}
Denoting
$$
Dw_n(z)=\sigma_n(z)+\tau_n(z),
$$
with
$\sigma_n(z)(q,q')=Dw_n(z)(q,q')\delta_{qq'}$, one computes from
\equ(recrelDw) the recursive relation
$$
\sigma_n(z)=\bigl(\tilde\sigma_{n-1}(\tilde z)+T_n(z)\bigr)
H_n(\tilde z),
\equation(recrelsn1)
$$
where
$$\eqalign{
H_n(\tilde z)&=\bigl(1-\Gamma_n\tilde\sigma_{n-1}(\tilde z)
\bigr)^{-1},\cr
T_n(z)(q,q')&=\bigl[\tau_n(z)\Gamma_n\tau_{n-1}(\tilde z)\bigr]
(q,q')\delta_{qq'}.\cr
}$$
Setting
$$
\RR_n(z)\equiv\tilde\sigma_{n-1}\bigl(H_n(\tilde z)-1\bigr),\quad
\TT_n(z)\equiv T_n(z)H_n(\tilde z),
$$
and using $\Delta_p\tilde\sigma_{n-1}=\Delta_p\sigma_{n-1}$ together
with the identity $\Delta_p\sigma_0=0$, one applies \equ(recrelsn1)
recursively to obtain
$$
\Delta_p\sigma_n(z)=\sum_{m=1}^{n}\Delta_p\bigl(
\RR_m(z_m)+\TT_m(z_m)\bigr),
\equation(recrelsn2)
$$
where $z_m=F_n^{m+1}(z)$, cf \equ(deffnm), with
$F_n^{n+1}\equiv\identity$.
Note that $\RR_m(z)$ is diagonal and can be rewritten as
$$
\RR_m(z)=\tilde\sigma_{m-1}(\tilde z)
\Gamma_m
\tilde\sigma_{m-1}(\tilde z)H_m(\tilde z).
\equation(proddiagonal)
$$
As shown below, each term in \equ(recrelsn2) is easily
seen to be of order
$\varepsilon^2|\omega\cdot p|$. Thus,
the main issue in obtaining \equ(Dpsigman) is to ensure that taking
the sum will deteriorate the bound only slightly.
Let us first consider the terms involving the quantities
$\Delta_p\TT_m$.
They are higher order terms, since $\TT_m$ is quadratic
in the off-diagonal
part $\tau_m$ which, as shown in \clm(delta1wn), are bounded
by powers of $r$. Indeed, as carried out in the
Appendix, one has for all $m=1,\dots,n$ and $z\in B_m$,
$$
||\hat P_m\Delta_p\TT_m(z)||^{(m)}_{s,s'}\leq
\varepsilon^2\eta^m|\omega\cdot p|,
\equation(bndonDpTTm)
$$
so that
$$
||\sum_{m=1}^n\hat P_n\Delta_p\TT_m(z)||^{(n)}_{s,s'}\leq
\sum_{m=1}^n||\hat P_m\Delta_p\TT_m(z)||^{(m)}_{s,s'}\leq
\varepsilon^2|\omega\cdot p|.
\equation(DpTTfinal)
$$
On the other hand,
the terms involving $\Delta_p\RR_m$ are not higher order terms.
Since
$$
\Delta_pH_m(\tilde z)=t_pH_m(\tilde z)\Bigl(\Delta_p\Gamma_m
t_p\tilde\sigma_{m-1}(\tilde z)
+\Gamma_m\Delta_p\tilde\sigma_{m-1}(\tilde z)
\Bigr)H_m(\tilde z),
$$
\equ(DeltapG) with $\sigma=s+\xi-\gamma$ and $n$ replaced by $m$
yields with the recursive bound \equ(Dpsigman)
$$
||\Delta_pH_m(\tilde z)||^{(m-2)}_{s}\leq\eta^{-m}
|\omega\cdot p|.
\equation(jj3)
$$
Thus, using in addition the recursive
bounds \equ(thdbnd) and \equ(Dpsigman), together with
$$
||H_m(\tilde z)-1||^{(m-1)}_{s}=
||\Gamma_m\tilde\sigma_{m-1}(\tilde y)H_m(\tilde z)||^{(m-1)}_{s}
\leq C\varepsilon,
$$
one obtains for all $m=1,\dots,n$ and $z\in B_m$,
$$
||\hat P_n\Delta_p\RR_m(z)||^{(n)}_{s,s'}\leq
||\hat P_m\Delta_p\RR_m(z)||^{(m)}_{s,s'}\leq
C\varepsilon^2|\omega\cdot p|,
\equation(app11)
$$
to be compared with \equ(bndonDpTTm).
However, one can actually show that
$$\eqalignno{
\Bigr\|\sum_{m=1}^n\hat P_n\Delta_p\RR_m(z)\Bigl\|^{(n)}_{s,s'}&\leq
\sup_{{\bf k}\in\II^n}\sup_{q\in S^n_{\bf k}}\sum_{m=1}^n
|\Delta_p\RR_m(z)(q)|_{s,s'}&\equat(bbbinter)\cr
&\leq C\varepsilon^2|\omega\cdot p|,
&\equat(bndonDpRRm)\cr
}$$
with another $n$-independent constant $C$.
Although \equ(app11) yields the a priori bound
$|\Delta_p\RR_m(z)(q)|_{s,s'}\leq C\varepsilon^2|\omega\cdot p|$
for all $q\in S^n_{\bf k}$ and ${\bf k}\in\II^n$,
\equ(bndonDpRRm) will follow from the fact that all but a finite
number of terms in \equ(bbbinter) are identically zero.
More precisely, there is
for all ${\bf k}\in\II^n$
a set $\ZZ^n_{\bf k}\subset\{1,\dots,n\}$ with $\#\ZZ^n_{\bf k}$
uniformly bounded in $n$ and ${\bf k}$ such that for all
$q\in S^n_{\bf k}$,
$$
|\Delta_p\RR_m(z)(q)|_{s,s'}\equiv0
\quad{\rm if}\quad m\not\in\ZZ^n_{\bf k}.
\equation(app2)
$$
This leads to \equ(bndonDpRRm) and concludes
the proof of bound \equ(Dpsigman), since \equ(recrelsn2),
\equ(DpTTfinal) and \equ(bndonDpRRm)
lead to \equ(Dpsigman) by taking $\varepsilon$ small
enough and by noting that $z_m\in B_m$ for all $z\in B_n$,
cf. \equ(fnmin).
Identity \equ(app2) for some finite set $\ZZ^n_{\bf k}$ follows from
the expression \equ(proddiagonal) for $\RR_m$ since by localization of
scales $\Gamma_m(q)=(A^{-1}Q_mP_{m-1})(q)=0$
for most $m\leq n$ if $q\in S^n_{\bf k}$.
More precisely, one computes that
$$
Q_m(q)P_{m-1}(q)=
\sum_{\tilde{\bf k}\in\II^m}\bigl(1-\chi^m_{\tilde{\bf k}}(q)\bigr)
\chi^{m-1}_{\tilde{\bf k}_{m-1}}(q)\PP^m_{\tilde{\bf k}},
$$
where the index $\tilde{\bf k}_{m-1}$
serves to denote the (unique) subspace
$\JJ^{m-1}_{\tilde{\bf k}_{m-1}}$
containing $\JJ^m_{\tilde{\bf k}}$.
Fix now some ${\bf k}\in\II^n$.
Then one has for all $q\in S^n_{{\bf k}}$
and all $m<n$,
$$
Q_m(q)P_{m-1}(q)=
\sum_{{\scriptstyle\tilde{\bf k}\in\II^m
\atop\scriptstyle\tilde{\bf k}\not={\bf k}_m}}
\chi^{m-1}_{\tilde{\bf k}_{m-1}}(q)\PP^m_{\tilde{\bf k}}
=\PP_{\JJ^{m-1}_{{\bf k}_{m-1}}\setminus\JJ^{m}_{{\bf k}_{m}}},
$$
since by construction $\chi^m_{{\bf k}_m}(q)=1$ for such $m$ and $q$.
Therefore, $Q_m(q)P_{m-1}(q)=0$ for all $q\in S^n_{\bf k}$ if $m<n$ is
such that $\JJ^{m}_{{\bf k}_{m}}=\JJ^{m-1}_{{\bf k}_{m-1}}$.
On the other hand, $\JJ^{m}_{{\bf k}_{m}}$
is a strict subspace of $\JJ^{m-1}_{{\bf k}_{m-1}}$ only if
$\#\CC^{m}_{{\bf k}_{m}}<\#\CC^{m-1}_{{\bf k}_{m-1}}$, i.e.,
if the eigenvalues contained in $\CC^{m-1}_{{\bf k}_{m-1}}$ have been
divided after perturbation by $a_{m-1}$ into two
(or more) clusters. But this can be true only for finitely many $m$
since the original eigenvalues $\mu_k$ are finitely many
times degenerate. Hence, there is an $L<\infty$
such that for all $n\geq1$ and all $1\leq m\leq n$, one has
$\hat P_n\RR_m(q)=0$,
except for some $m_1,\dots,m_L$. Since the same is true of
$\hat P_nt_p\RR_m(q)$ provided that $p$ satisfies
$|\omega\cdot p|<\eta^{n-1}/16$, \equ(app2) follows.
\qed

\subsection The Cluster Decomposition

We now check that point (d) of \clm(itbounds) holds. First,
\equ(deltan), \equ(defGn) and \equ(subinv) lead to,
for ${\bf k}=(k,\cdot)\in\II^n$,
$$
|a_n\PP^n_{\bf k}|_{\LL(\JJ^n_{\bf k})}\leq
3k^{\gamma-\xi}\varepsilon\eta^{n-1},
\equation(ee1)
$$
which, since $\mu_k^2\geq ck^{2\gamma}$ by hypothesis (H1), implies
that $\mu^2+\sum_{m=0}^na_m\equiv\tilde\mu_{n+1}^2$
is positive definite for
$\varepsilon=\varepsilon(c,\eta)$ small enough. Next, it follows from
$a_n$ being hermitian that $\sigma(\tilde \mu_{n+1})\subset\real^+$.
Furthermore, using \equ(subinv) and the fact that $\JJ^n_{\bf k}$ is
by definition an invariant subspace for $\tilde\mu_n$, one infers from
$\mu_k\geq ck^\gamma$,
the asymptotic \equ(asymp) for $\tilde\mu_n$, and the estimate
\equ(ee1), that
$$
|a_n\tilde\mu_n^{-1}\PP^n_{\bf k}|_{\LL(\JJ^n_{\bf k})}
\leq3c^{-1}k^{-\xi}\varepsilon\eta^{n-1}.
$$
Therefore, denoting by $\PP_k$ the projector
onto the $k^{\rm th}$ component of
$\hat\RR^\infty=\bigoplus_{k\geq1}\complex^{d_k}$,
one obtains
$$
\tilde\mu_{n+1}\PP_{k}=
\bigl[\tilde\mu_n^2+a_n\bigr]^\onehalf\PP_{k}
=\tilde\mu_n\PP_{k}+\OO(k^{-\xi}\varepsilon\eta^{n-1}),
\equation(ee2)
$$
which, since $\mu\PP_k=\mu_k\identity_{d_k}$, implies by recursion
that
$$
\tilde\mu_{n+1}\PP_{k}=\mu_k\identity_{d_k}
+\OO(\varepsilon k^{-\xi}).
$$
Hence,
the
asymptotic \equ(asymp) holds, where for each $k\geq1$
the sequence of clusters $\CC^{n+1}_{k,i}$, $i=1,\dots,M_{k}^{n+1}$,
forms a partition
of the component $\sigma(\tilde\mu_{n+1}\PP_k)$
satisfying
$d(\CC^{n+1}_{k,i},\CC^{n+1}_{k,j})>\eta^{n+1}$ for $i\not=j$.
This partition is unique if
$M_k^{n+1}$ is required to be maximal.
Furthermore,
it follows from \equ(asymptotics2) and \equ(asymptotics3) in
(H1) that for $\varepsilon=\varepsilon(c)$ small enough,
the components $\sigma(\tilde\mu_{n+1}\PP_k)$ are well separated.
Therefore, the sets $S^{n+1}_{\bf k}$, ${\bf k}\in\II^{n+1}$, defined
according to \equ(defsin) are pairwise disjoint.
Next, \equ(ee2) and the gap condition \equ(gapn2) with $n+1$ replaced
by $n$ imply that for
$\varepsilon=\varepsilon(c,\eta)$ small enough, every cluster
$\CC^{n+1}_{k,i}$ is composed of perturbed eigenvalues belonging to a
unique $\CC^n_{k,j^{n+1}_i}$. The distance between these two clusters
is at most of order $\OO(k^{-\xi}\varepsilon\eta^{n-1})$, so that
\equ(convclust) follows for $n+1$ by induction.
In order to iterate \equ(PkiPlj), we note that
by definition, $\JJ^{n+1}_{\bf k}$ is the eigenspace of
$\tilde\mu_{n+1}$ associated with $\CC^{n+1}_{\bf k}$,
${\bf k}\in\II^{n+1}$, and that
every $\JJ^n_{{\bf k}'}$,
${\bf k}'\in\II^{n}$, is also an invariant
subspace for $\tilde\mu_{n+1}$ by \equ(subinv).
Therefore, each $\JJ^{n+1}_{k,i}$ is contained in a unique
$\JJ^n_{k,j^{n+1}_i}$, namely, the eigenspace associated with
$\CC^n_{k,j^{n+1}_i}$. Finally, we check that \equ(proppn) iterates.
This is a simple consequence of \equ(PkiPlj) and
$S^{n+1}_{k,i}\subset S^n_{k,j^{n+1}_i}$,
the latter following from \equ(ee2)
for $\varepsilon$ small enough.

\section Measure Estimate

In this section, we prove \clm(measureest), namely, that
$\Omega^*(K)=\bigcap_{n\geq1}\Omega_n(K)$ satisfies
$$
\lim_{K\rightarrow0}{\rm meas}(\Omega\setminus\Omega^*(K))=0,
\equation(measest)
$$
for all bounded $\Omega\subset\real^d$.
The strategy is standard and consists in studying the
complementary sets of $\Omega_n(K)$.
For $n\geq1$, $b>0$, and $q\in\integer^d$, let us define
$$
\Sigma^n_{q,b}\equiv\Bigl(\bigcup_{\bf k\in\II^n}
\Sigma^{n;\bf k}_{q,b}\Bigr)
\cup\Bigl(\bigcup_{{\bf k,k'}\in\II^n}
\Sigma^{n;{\bf k,k'}}_{q,b}\Bigr),
$$
where
$$\eqalign{
\Sigma^{n;\bf k}_{q,b}&=\{\omega\in\real^d\ |\
d(|\omega\cdot q|,\CC^n_{\bf k})\leq b\},\cr
\Sigma^{n;\bf k,k'}_{q,b}&=\{\omega\in\real^d\ |\
d(|\omega\cdot q|,|\CC^n_{\bf k}\pm\CC^n_{\bf k'}|)\leq b\}.\cr
}$$
Next, with
$$
\ZZ_n\equiv\{q\in\integer^d\ |\
K^{1\over\nu}\eta^{-{n-1\over\nu}}\leq|q|<
K^{1\over\nu}\eta^{-{n\over\nu}}\},
$$
and
$$
\Sigma^*(K)\equiv\bigcup_{n\geq1}\bigcup_{q\in\ZZ_n}
\Sigma^n_{q,2K|q|^{-\nu}},
$$
one shows first, that for all bounded $\Omega\subset\real^d$,
$$
{\rm meas}\bigl(\Omega\cap\Sigma^*(K)\bigr)\leq
C_\Omega\,K^{\xi\over\xi+1},
\equation(bndmeas)
$$
for some constant $C_\Omega$ depending on $\Omega$ only,
and, second, that
$$
\bigl[\Sigma^*(K)\bigr]^c\subseteq
\Omega^*(K).
\equation(inclusion)
$$
Obviously, \equ(measest) follows from \equ(bndmeas) and
\equ(inclusion).
Below, $C_\Omega$ will denote a generic constant that
may change from place to place but depends on $\Omega$ only.

Let us start with the bound \equ(bndmeas). One has
$$
{\rm meas}\bigl(\Omega\cap\Sigma^*(K)\bigr)
\leq\sum_{n\geq1}\sum_{q\in\ZZ_n}\bigl(T^{n}_{q,2K|q|^{-\nu}}
+\hat T^{n}_{q,2K|q|^{-\nu}}\bigr),
\equation(decomp1)
$$
where
$$
T^{n}_{q,b}=\sum_{{\bf k}\in\II^n}{\rm meas}
\bigl(\Omega\cap\Sigma^{n;\bf k}_{q,b}\bigr)\,,\
\hat T^n_{q,b}={\rm meas}\Bigl(\Omega\cap
\bigcup_{{\bf k,k'}\in\II^n}\Sigma^{n;\bf k,k'}_{q,b}\Bigr).
\equation(T1T2)
$$
To treat the terms on the right hand side of \equ(decomp1) involving
the quantities $T^{n}_{q,b}$, we first
use \equ(bndabsIin) to estimate,
$$
{\rm meas}\bigl(\Omega\cap\Sigma^{n;{\bf k}}_{q,b}\bigr)\leq
C_\Omega(b+\bar d\eta^n).
$$
Next, we note
that the asymptotic behavior of the clusters $\CC^n_{\bf k}$,
cf. \equ(asymptotics1) and \equ(asymp), implies that
$\Omega\cap\Sigma^{n;\bf k}_{q,b}$ is empty if ${\bf k}=(k,\cdot)$
satisfies $k\geq C_\Omega|q|$ for some constant $C_\Omega$.
Hence, since the number of indices ${\bf k}$ of
the form $(k,\cdot)$ is uniformly bounded in $k$,
the number of terms which are non-zero in the sum
defining $T^{n}_{q,b}$
is proportional to $|q|$, and one obtains the estimate
$T^{n}_{q,b}\leq C_{\Omega}|q|(b+\bar d\eta^n)$.
Finally, the fact that
$q\in\ZZ_n$ satisfies
$\eta^n\leq K|q|^{-\nu}$ leads to
$$
\sum_{n\geq1}\sum_{q\in\ZZ_n}T^{n}_{q,2K|q|^{-\nu}}
\leq C_\Omega\bigl(2K+\bar dK\bigr)
\sum_{q\in\integer^d}|q|^{1-\nu}
\leq C_\Omega\, K,
\equation(f1)
$$
for $\nu=\nu(d)$ large enough.
To treat the remaining terms in \equ(decomp1),
we first note that, as above,
$$
{\rm meas}\bigl(\Omega\cap\Sigma^{n;{\bf k,k'}}_{q,b}\bigr)\leq
C_\Omega(b+2\bar d\eta^n).
\equation(bbb1)
$$
Next, one distinguishes the
cases $\gamma=1$ and $\gamma>1$. If $\gamma>1$, then for $k'>k$
the inequality $k'^\gamma-k^\gamma>k'^{\gamma-1}$ and the asymptotic
\equ(asymptotics2) imply that
$\Omega\cap\Sigma^{n;{\bf k,k'}}_{q,b}$ is empty for
${\bf k}=(k,\cdot)$ and ${\bf k'}=(k',\cdot)$ such that
$k'\geq C_\Omega|q|^{1/(\gamma-1)}\equiv k_q$.
Furthermore, it follows from \equ(asymp) that for
$k_b=b^{-{1\over\xi+1}}$,
$$
{\rm meas}\Bigl(\bigcup_{\scriptstyle k>Ck_b\atop\scriptstyle i,j}
\Sigma^{n;(k,i),(k,j)}_{q,b}\Bigr)\leq Ck_b^{-\xi}.
$$
Therefore, one obtains with \equ(bbb1)
$$\eqalign{
\hat T^n_{q,b}&\leq Cb^{{\xi\over\xi+1}}+
\sum_{k=1}^{Ck_b}{\rm meas}
(\Omega\cap\Sigma^{n;(k,i),(k,j)}_{q,b})
+\sum_{\scriptstyle k'=2\atop\scriptstyle k<k'}^{k_q}
{\rm meas}(\Omega\cap\Sigma^{n;(k,i),(k',j)}_{q,b})\cr
&\leq Cb^{{\xi\over\xi+1}}+C_\Omega(b+2\bar d\eta^n)
\Bigl(b^{-{1\over\xi+1}}+|q|^{2\over \gamma-1}\Bigr).\cr
}$$
This finally leads to, using again that
$\eta^n\leq K|q|^{-\nu}$
for $q\in\ZZ_n$,
$$
\sum_{n\geq1}\sum_{q\in\ZZ_n}\hat T^{n}_{q,2K|q|^{-\nu}}\leq
CK^{\xi\over1+\xi}\sum_{q\in\integer^d}|q|^{-\nu{\xi\over1+\xi}}+
C_\Omega K\sum_{q\in\integer^d}|q|^{{2\over \gamma-1}-\nu}
\leq C_\Omega K^{\xi\over1+\xi},
\equation(f2)
$$
for $\xi>0$ and $\nu=\nu(d,\xi)$ large enough.
We now consider the case $\gamma=1$. From \equ(asymp)
and the asymptotic
behavior \equ(asymptotics3), it follows first that
$\Omega\cap\Sigma^{n;{\bf k,k'}}_{q,b}$ is empty for
${\bf k}=(k,\cdot)$ and ${\bf k'}=(k',\cdot)$ with
$k'-k=l\geq C|q|$, and second that for all $l\geq0$
$$
{\rm meas}\Bigl(\bigcup_{\scriptstyle k>Ck_b\atop\scriptstyle i,j}
\Sigma^{n;(k,i),(k+l,j)}_{q,b}\Bigr)\leq Ck_b^{-\xi},
$$
where $k_b=b^{-{1\over\xi+1}}$.
Therefore, \equ(bbb1) leads to
$$
\hat T^n_{q,b}\leq C|q|b^{{\xi\over\xi+1}}+
C_\Omega b^{-{1\over\xi+1}}|q|(b+2\bar d\eta^n),
$$
and one finally obtains for $\nu=\nu(d,\xi)$ large enough,
$$
\sum_{n\geq1}\sum_{q\in\ZZ_n}\hat T^{n}_{q,2K|q|^{-\nu}}\leq
C_\Omega K^{\xi\over1+\xi}\sum_{q\in\integer^d}
|q|^{1-\nu{\xi\over1+\xi}}
\leq C_\Omega K^{\xi\over1+\xi}.
\equation(f3)
$$
Inserting \equ(f1) and \equ(f3) into \equ(decomp1)
yields \equ(bndmeas).

We now check that \equ(inclusion) holds. If
$\omega\not\in\Sigma^*(K)$, then the following is true for all
$n\geq1$, $q\in\ZZ_n$ and ${\bf k,k'}\in\II^n$,
$$\eqalignno{
&d(|\omega\cdot q|,\CC^n_{\bf k})>2K|q|^{-\nu},&\equat(loc1)\cr
&d(|\omega\cdot q|,|\CC^n_{\bf k}\pm\CC^n_{\bf k'}|)>2K|q|^{-\nu}.
&\equat(loc2)\cr
}$$
Next, we verify that for such $\omega$, this implies that
bounds \equ(loc1) and \equ(loc2)
hold for all
$q\in\bigcup_{m=1}^n\ZZ_m$ provided one replaces the constant
$2K$ on the right hand side by $K$.
This in turn implies that
$\omega\in\Omega_n(K)$ for all $n\geq1$, so that
$\omega\in\Omega^*(K)$.
Let $m<n$ and fix some ${\bf k}\in\II^n$.
Then, recalling \equ(convclust), namely that
there is at least one
${\bf k'}\in\II^m$
for which
$$
\sup_{x\in I^n_{\bf k}}\inf_{y\in I^m_{\bf k'}}d(x,y)\leq\eta^{m+1},
$$
and since, on the other hand,
$\eta^m<K|q|^{-\nu}$ whenever $q\in\ZZ_m$,
one infers from
\equ(loc1) with $n$ replaced by $m$ that for $q\in\ZZ_m$ and $\eta<1$,
$$\eqalignno{
d(|\omega\cdot q|,\CC^n_{\bf k})
&\geq\bigl|d(|\omega\cdot q|,\CC^m_{\bf k'})-\eta^{m+1}\bigr|&\cr
&\geq(2K-\eta K)|q|^{-\nu}&\cr
&>K|q|^{-\nu}.&\equat(bbb3)\cr
}$$
Since \equ(bbb3) holds for all
$q\in\ZZ_m$,
$1\leq m\leq n$,
one concludes that
$d(|\omega\cdot q|,\CC^n_{\bf k})>K|q|^{-\nu}$ whenever
$0<|q|<K\eta^{-n/\nu}$.
In a similar way, one derives an identical lower bound on
$d(|\omega\cdot q|,|\CC^n_{\bf k}\pm\CC^n_{\bf k'}|)$,
thus achieving the
proof of \equ(inclusion) and \equ(measest).

\section Proof of \clm(mainthm)

Defining
$z_n\equiv F_n(0)$,
we now show that $z_n$ converges in $h_s$, as $n\rightarrow\infty$,
to a function $z$ whose Fourier transform is real analytic and
provides a solution of equation \equ(sys2).
Using $F_n(0)=F_{n-1}(R_n(0))$, cf. \equ(deffn),
one computes that
$$
z_n-z_{n-1}=\delta_1F_{n-1}\bigl(R_n(0)\bigr).
$$
According to \equ(equwtilde2), $R_n(0)=H_n\Gamma_nw_{n-1}(0)+u(0)$,
so that \equ(fstbnd),
\equ(bndinvan), \equ(boundonh), \equ(ballBB) and the identity
$\Gamma_n=\Gamma_n\hat P_{n-1}$ lead to
$$
||R_n(0)||_{h^{n-1}_s}\leq\eta^{-n}r^{2(n-1)}.
$$
Therefore,
since, $F_{n-1}\in\AA_{n-1}=H^\infty(B_{n-1},h_{s'})$,
one can apply \equ(Cauchy3)
to $\delta_1F_{n-1}$ with $\gamma=\eta^{-n}r^{n-2}$ to obtain
$$
||z_n-z_{n-1}||_s\leq C\eta^{-n}r^{n-2}|||F_{n-1}|||_{\AA_{n-1}},
$$
and the convergence of $z_n$ in $h_s$ follows from the uniform bound
\equ(unifFn) by taking $r=r(\eta)$ small enough.
Bound \equ(unifFn) also implies $||z_\beta||\leq\varepsilon$
uniformly in the strip
$|\Im\beta|<\alpha'=\alpha\prod_{n=2}^\infty(1-n^{-2})$.
This yields the
pointwise estimate
$$
|z(q)|\leq\varepsilon e^{-\alpha'|q|},
$$
and, consequently,
ensures the real analyticity of the Fourier transform of $z$.

In order to prove that the limit $z$ solves equation \equ(fp1),
namely, $\sK_0z=w_0(z)$, we will
show below that
$$
\sK_0z_n=Q_nw_0(z_n)+A_{<n}P_nR_n(0).
\equation(Ayn1)
$$
where one has defined
$A_{<m}\equiv\sum_{k=0}^{m-1}A_k$ for $m=1\dots,n$.
Since it follows from \equ(tildeyinbn) and \equ(deltan) that
$$
||A_kR_n(0)||_{s'}\leq C\varepsilon\eta^{k-1}r^n,
$$
the second term in the right hand side of \equ(Ayn1)
converges to zero in $h_{s'}$ as $n\rightarrow\infty$.
Moreover,
$\lim_{n\rightarrow\infty}Q_n=\identity$ for each
$\omega\in\Omega^*$, and
since $w_0$ is analytic, one can take the
$n\rightarrow\infty$ limit of equation \equ(Ayn1) to
conclude that $z$ solves \equ(fp1).
It thus remains to check that identity \equ(Ayn1) holds.
We will use the following relations
$$\eqalignno{
&z_n=\sum_{m=1}^n R_m(F_n^{m+1}(0)),&\equat(id1)\cr
&w_m(F_n^{m+1}(0))=
w_0(z_n)-\sum_{k=0}^{m-1}A_kF_n^{k+1}(0),&\equat(id2)\cr
}$$
where $F_n^m$ is defined by \equ(deffnm) for $m\leq n$, whereas
$F_n^{n+1}\equiv\identity$.
The first relation simply follows from $z_n=F_n^1(0)$ by
using recursively the definitions \equ(deftildeRn) and \equ(deffnm).
The second relation is obtained by using \equ(defwn) and
\equ(defwtilden) to get
$$
w_m(F_n^{m+1}(0))=
w_{m-1}\bigr(F_n^m(0)\bigl)-A_{m-1}F_n^m(0),
$$
which one applies recursively.
Next,
it follows from \equ(id1) that
$$
\sK_0z_n=\sum_{m=1}^n\sK_0R_m(F_n^{m+1}(0)),
\equation(Ayn0)
$$
whereas \equ(id2) and \equ(defwn)
imply, since
$R_m$ solves equation \equ(equrn) with $n$ replaced by $m$,
$$\eqalignno{
\sK_mR_m(F_n^{m+1}(0))&=Q_mP_{m-1}\tilde w_{m-1}(F_n^m(0))&\cr
&=Q_mP_{m-1}\Bigl(w_0(z_n)-\sum_{k=0}^{m-1}A_kF_n^{k+1}(0)\Bigr),
&\equat(AmRm)\cr
}$$
where, for $m=1$, one denotes $P_0\equiv\identity$.
Therefore, since $\sum_{m=1}^nQ_mP_{m-1}=Q_n$ and
$\sK_0=\sK_m+\sum_{k=0}^{m-1}P_kA_k$,
\equ(Ayn0) and \equ(AmRm) yield
$$
\sK_0z_n=Q_nw_0(z_n)+\sum_{m=1}^nT_m,
\equation(Ayn2)
$$
where $T_m$ is given by
$$
T_m=\sum_{k=0}^{m-1}\Bigl(P_kA_kR_m(F_n^{m+1}(0))-
Q_mP_{m-1}A_kF_n^{k+1}(0)\Bigr).
\equation(defTm)
$$
If $P_k$ and $Q_k$ were true projectors, namely, if the scales were
defined in terms of sharp cut-off functions, cf. \equ(defPn), then a
straightforward calculation would show that $T_m\equiv0$.
Nevertheless,
although none of the quantities $T_m$ are zero, we
check that
$$
T_1=A_{<1}\hat R_1\quad{\rm and}\quad
T_m=A_{<m}\hat R_m-A_{<m-1}\hat R_{m-1}\quad{\rm for}
\quad m=2,\dots,n,
\equation(Tdiff)
$$
where $\hat R_n=P_nR_n(0)$ and for $m=1,\dots,n-1$,
$$
\hat R_m=P_{m}R_{m}(F_{n}^{m+1}(0))-Q_{m}R_{m+1}(F_{n}^{m+2}(0)).
$$
Thus, one is left with the small correction term
$\sum_{m=1}^nT_m=A_{<n}P_nR_n(0)$ as claimed in \equ(Ayn1).
To show \equ(Tdiff), we note that $A_k$ commutes with $P_m$ and $Q_m$
for all $k,m$. Furthermore, one easily checks that
$Q_mR_{m-1}=R_{m-1}$, $P_{m-1}R_{m+1}=R_{m+1}$, and
$Q_mP_{m-1}R_l=0$ if $l\not=m-1,m$, or $m+1$.
Hence, using in addition
$$
F_n^{k+1}(0)=\sum_{l=k+1}^n R_l(F_n^{l+1}(0)),
$$
one obtains, decomposing the expression for $T_m$ \equ(defTm) as
$T_m=T_m^{(1)}-T_m^{(2)}$,
$$\eqalign{
T_m^{(1)}&=A_{<m-1}R_m+A_{m-1}P_{m-1}R_m\,,\cr
T_m^{(2)}&=A_{<m-1}P_{m-1}R_{m-1}+A_{<m}\bigl(Q_mP_{m-1}
R_m+Q_mR_{m+1}\bigr),\cr
}$$
where $A_{<m-1}\equiv0$ for $m=1$ and the last term in $T_m^{(2)}$ is
absent for $m=n$.
Substracting $T^{(2)}_m$ from $T^{(1)}_m$ finally leads to
\equ(Tdiff) by using
the identities
$(1-Q_m)P_{m-1}=P_m$ and $1-Q_mP_{m-1}=Q_{m-1}+P_m$.
This completes the proof that
$z=\lim_{n\rightarrow\infty}z_n$ solves \equ(fp1).

The resulting solution $Z=Z(\lambda,\Phi,J)$
of equation \equ(sys2) depends analytically
on $\lambda$ for $|\lambda|<\lambda_0$ and vanishes for
$\lambda=0$. Its uniqueness follows from the fact that equation
\equ(sys2) completely determine the coefficients of the Taylor
expansion of its solution in powers of $\lambda$. Furthermore,
recall that $\Phi$ and $J$ are parameters in $w_n$, the latter being
analytic in $(\Phi,J)\in\sO_\sB$. Thus, $Z$ is also analytic in
$(\Phi,J)\in\sO_\sB$.

These properties can be used, together with \clm(KAM),
to conclude the proof of \clm(mainthm), namely, to check
that equation \equ(eqmot3) has a unique solution $\TT=(\Phi,J,Z)$, up
to translation \equ(transinv), analytic in $\lambda$ and vanishing for
$\lambda=0$.
To this end, introducing the variable $Y=(\Phi,J)$,
we denote by $Y_s(\lambda,Z)$ the solution of \equ(sys1) and by
$Z_s(\lambda,Y)$ the solution of \equ(sys2). Then, the solution
$\TT(\lambda)$ of \equ(eqmot3) is given by
$\TT=(Y_s(\lambda,\ZZ),\ZZ)$
where $\ZZ=\ZZ(\lambda)$ solves the functional fixed point equation
$$
\ZZ=Z_s(\lambda,Y_s(\lambda,\ZZ))\equiv\FF(\ZZ,\lambda).
\equation(fpe)
$$
To solve \equ(fpe) for $\ZZ(\lambda)$,
we use the implicit function theorem.
We first note that by \clm(KAM), $Y_s(\lambda,Z)$
is well defined in $\sB$ for
$|\lambda|<\lambda_1$ and $Z$ in a small enough neighborhood of the
origin $\sO_s\subset h_s$, with $Y_s(\lambda,Z)|_{\lambda=0}=0$ for
all $Z\in\sO_s$.
Hence, there is a $\lambda_2>0$ small enough such that
$Y_s(\lambda,Z)\in\sO_\sB$ for $|\lambda|<\lambda_2$ and $Z\in\sO_s$.
It thus follows from the previous discussion that
$\FF$ is analytic in $|\lambda|<\lambda_2$ and $\ZZ\in\sO_s$
with $\FF(\lambda,\ZZ)\in h_s$ and
$\FF(\lambda,\ZZ)|_{\lambda=0}=0$ for all $\ZZ\in\sO_s$.
One infers, in particular, that the solution
of \equ(fpe) at $\lambda=0$ is given by
$\ZZ(\lambda)|_{\lambda=0}=0$.
Next, one computes
$$
D_\ZZ\FF(\ZZ,\lambda)=D_Y Z_s(\lambda,Y_s(\lambda,\ZZ))
D_\ZZ Y_s(\lambda,\ZZ).
$$
Since $Y_s(\lambda,Z)|_{\lambda=0}=0$ for all
$Z\in\sO_s$ and $Z_s(\lambda,Y)|_{\lambda=0}=0$ for all
$Y\in\sO_\sB$, it follows that
$D_\ZZ Y_s(\lambda,\ZZ)|_{\lambda=0}=
D_Y Z_s(\lambda,Y_s(\lambda,\ZZ))|_{\lambda=0}=0$ for $\ZZ\in\sO_s$,
which, in turn, implies
$$
D_\ZZ\FF(\lambda,\ZZ)|_{\scriptstyle\lambda=0\atop
\scriptstyle\ZZ=0}=0.
$$
Therefore,
the existence for all $|\lambda|<\lambda_2$ of a unique
$\ZZ(\lambda)\in\sO_s$ solving \equ(fpe) follows by the implicit
function theorem.

\sectionnonr Appendix

\noindent
{\bf Proof of \clm(intermediary).}

\noindent
We first consider the estimate \equ(bndinvan). Since $\Gamma_n$ has a
diagonal kernel, one has
$$
||\Gamma_n||_{\sigma,\sigma+\gamma}=
\sup_{q\in\integer^d}|\sK_n^{-1}(q)Q_n(q)
P_{n-1}(q)|_{\sigma,\sigma+\gamma}\ ,
$$
and using \equ(PkiPlj) one easily computes that
$$
\Gamma_n(q)=
\sum_{{\bf k}\in\II^n}\sK_n^{-1}(q)\hat\chi^n_{\bf k}(q)\PP^n_{\bf k},
\equation(expGn)
$$
whith
$$
\hat\chi^n_{\bf k}(q)=
\bigl(1-\chi^n_{\bf k}(\omega\cdot q)\bigr)
\chi^{n-1}_{{\bf k}_{n-1}}(\omega\cdot q),
\equation(defhatchi)
$$
where ${\bf k}_{n-1}\in\II^{n-1}$ labels
the unique subspace
$\JJ^{n-1}_{{\bf k}_{n-1}}$
containing $\JJ^n_{\bf k}$.
Note that $\hat\chi^n_{\bf k}(q)\not=0$ only for
$q$ in the set
$$
\hat S^n_{\bf k}=\{q\in\integer^d\ |\
d(|\omega\cdot q|,\CC^n_{\bf k})\geq\oneeighth\eta^n\ {\rm and}\
d(|\omega\cdot q|,\CC^{n-1}_{{\bf k}_{n-1}})\leq\onefourth\eta^{n-1}
\}.
\equation(defhatSin)
$$
Although the sets $\hat S^n_{\bf k}$ are not pairwise disjoint,
$\hat S^n_{\bf k}\cap\hat S^n_{\bf k'}\not=\emptyset$ only if
${\bf k}_{n-1}={\bf k'}_{n-1}$.
Since for ${\bf k}=(k,i)$
this happens only if ${\bf k'}=(k,j)$, and
since the original frequencies $\mu_k$ are by assumption finitely
many times degenerate (uniformly in $k$), there are for all
${\bf k}\in\II^n$
no more than $\bar d$ ${\bf k'}$ such that
${\bf k'}_{n-1}={\bf k}_{n-1}$.
Therefore, one obtains, with
$0\leq\hat\chi^n_{\bf k}\leq1$,
$$
||\Gamma_n||_{\sigma,\sigma+\gamma}\leq \bar d\sup_{{(k,i)}\in\II^n}
\sup_{q\in\hat S^n_{(k,i)}}k^\gamma|\sK_n^{-1}(q)\PP^n_{k,i}|.
\equation(bndG1)
$$
Let us now fix some ${\bf k}=(k,i)\in\II^n$ and $q\in\hat S^n_{\bf k}$
with $\omega\cdot q>0$.
Thus, $P_m(q)=\PP^m_{{\bf k}_m}$ for all $0\leq m\leq n-2$, whereas
$P_{n-1}(q)=\chi^{n-1}_{{\bf k}_{n-1}}(q)\PP^{n-1}_{{\bf k}_{n-1}}$.
This in turn imply,
since $\sK_n=\sK_0-\sum_{m=0}^{n-1}A_mP_m$ and $q\in\QQ^+_\omega$,
that $\sK_n(q)$ can be rewritten as
$$
\sK_n(q)=\hat \sK_n(q)+\hat\mu_n(q),
\equation(hatAhatmu)
$$
where
$$\eqalignno{
\hat \sK_n(q)&=|\omega\cdot q|^2-\mu^2-\sum_{m=0}^{n-1}
a_m\PP^m_{{\bf k}_m},
&\equat(defhatAn)\cr
\hat\mu_n(q)&=\bigl(1-\chi^{n-1}_{{\bf k}_{n-1}}(\omega\cdot q)\bigr)
a_{n-1}\PP^{n-1}_{{\bf k}_{n-1}}.
&\equat(defhatmun)\cr
}$$
Note that $\hat \sK_n$ is defined in such a way that
$(|\omega\cdot q|^2-\hat \sK_n(q))\PP^n_{\bf k}=\tilde\mu_n^2$,
cf. \equ(defmun).
One thus infers from the
definition of $\JJ^n_{\bf k}$ that $\JJ^n_{\bf k}$
is an invariant subspace of $\hat \sK_n(q)$.
Therefore, since the spectrum of
$(|\omega\cdot q|-\tilde\mu_n)\PP^n_{\bf k}$
is bounded away from zero by $\eta^n/8$, one
concludes from the identity
$|\omega\cdot q|^2-\tilde\mu_n^2=(|\omega\cdot q|-\tilde\mu_n)
(|\omega\cdot q|+\tilde\mu_n)$ and the asymptotic behavior
\equ(asymptotics1), \equ(asymp), that
$$
|\hat \sK_n^{-1}(q)\PP^n_{\bf k}|\leq Ck^{-\gamma}\eta^{-n}.
\equation(l32b1)
$$
On the other hand, denoting
$f(q)=1-\chi^{n-1}_{{\bf k}_{n-1}}(\omega\cdot q)$,
we compute that
$$
\hat \sK_n^{-1}(q)\hat\mu_n(q)=f(q)
\bigl(|\omega\cdot q|^2-\tilde\mu_n^2\bigr)^{-1}
\PP^{n-1}_{{\bf k}_{n-1}}a_{n-1}.
$$
Since $f(q)=0$ whenever
$d(|\omega\cdot q|,\CC^n_{\bf k'})\leq\eta^{n}$ for all
${\bf k'}$ such that
$\JJ^n_{\bf k'}\subset \JJ^{n-1}_{{\bf k}_{n-1}}$,
and since \equ(deltan)
(with $n$ replaced by $n-1$) implies
$|a_{n-1}|\leq3\varepsilon k^{\gamma-\xi}\eta^{n-2}$,
one estimates for
$\varepsilon$ small enough that
$\bigl|\hat \sK_n^{-1}(q)\hat\mu_n(q)\bigr|\leq k^{-\xi}/4\leq1/4$,
which leads to
$$
\bigl|\bigl(1+\hat \sK_n^{-1}(q)\hat\mu_n(q)\bigr)^{-1}\bigr|\leq2.
\equation(l32b2)
$$
Bound \equ(bndinvan) finally follows from \equ(bndG1)
by applying \equ(l32b1) and
\equ(l32b2) to
$$
\sK_n^{-1}(q)\PP^n_{\bf k}=
\bigl(1+\hat\sK_n^{-1}(q)\hat\mu_n(q)\bigr)^{-1}
\hat \sK_n^{-1}(q)\PP^n_{\bf k},
$$
and by noting that for $q\in\hat S^n_{\bf k}\cap\QQ^-_\omega$, the
previous analysis must be carried out with $\overline{a_m}$ instead of
$a_m$, $m=0,\dots,n-1$, and leads to identical bounds since $a_m$
being hermitian implies
$\sigma\bigr(\,\overline{\tilde\mu_n}\,\bigl)=\sigma(\tilde\mu_n)$.

To conclude the proof of \clm(intermediary), it remains to check
bound \equ(DeltapG).
If $p\in\integer^d$ is such that
$|\omega\cdot p|\geq\eta^{n+1}$, one can estimate
$$
||\Delta_p\Gamma_n||_{\sigma,\sigma+\gamma}
\leq2||\Gamma_n||_{\sigma,\sigma+\gamma}
\leq2\eta^{-n-1}||\Gamma_n||_{\sigma,\sigma+\gamma}\,|\omega\cdot p|,
$$
which, with \equ(bndinvan), leads to \equ(DeltapG)
for some other constant $C$.
Let us assume now that $|\omega\cdot p|<\eta^{n+1}$.
One computes from \equ(expGn) that
$$\eqalignno{
\Delta_p\Gamma_n(q)&=
\sum_{{\bf k}\in\II^n}\Bigl(\sK_n^{-1}(q+p)\hat\chi^n_{\bf k}(q+p)-
\sK_n^{-1}(q)\hat\chi^n_{\bf k}(q)\Bigr)\PP^n_{\bf k}&\cr
&=\sum_{{\bf k}\in\II^n}\Delta_p\sK_n^{-1}(q)
\PP^n_{\bf k}t_p\hat\chi^n_{\bf k}(q)
+\sum_{{\bf k}\in\II^n}\sK_n^{-1}(q)\PP^n_{\bf k}
\Delta_p\hat\chi^n_{\bf k}(q).
&\equat(l32b3)\cr
}$$
We now fix some ${\bf k}=(k,i)\in\II^n$ and
start by considering the second sum on the
right hand side of \equ(l32b3).
Since $p$ is such that $|\omega\cdot p|<\eta^{n+1}$,
$\Delta_p\hat\chi^n_{\bf k}(q)$ is non zero only for $q$ in a set
$\tilde S^n_{\bf k}$ that satisfies, with respect to the cluster
$\CC^n_{\bf k}$,
similar gap condition as $\hat S^n_{\bf k}$.
Therefore, the bounds derived
previously imply that
$|\sK_n^{-1}(q)\PP^n_{\bf k}|\leq Ck^{-\gamma}\eta^{-n}$ for
$q\in\tilde S^n_{\bf k}$, and one concludes by noting that
$$
|\Delta_p\hat\chi^n_{\bf k}(q)|\leq C\eta^{-n}|\omega\cdot p|,
$$
for all $q\in\integer^d$.
We now consider the first sum on the right hand side of \equ(l32b3).
Let us fix $q$ satisfying
$t_p\hat\chi^n_{\bf k}(q)\not=0$.
Using the same notation, we decompose $\sK_n$
as in \equ(hatAhatmu) and express
$$
\Delta_p\sK_n^{-1}(q)=\Bigr[
\bigl(1+t_p(\hat \sK_n^{-1}\hat\mu_n)\bigr)^{-1}
\Delta_p\hat \sK_n^{-1}
\bigl(1+\hat\mu_n\hat \sK_n^{-1}\bigr)^{-1}\Bigl](q).
$$
Bound \equ(l32b2) implies that
$\bigl|\bigl(1+t_p(\hat\sK_n^{-1}\hat\mu_n)(q)\bigr)^{-1}
\bigr|\leq2$.
Since $|\omega\cdot p|<\eta^{n+1}$, it follows that
$\bigl|\bigl(1+\hat \sK_n^{-1}\hat\mu_n(q)\bigr)^{-1}\bigr|$
satisfies a similar bound for $q$ with
$t_p\hat\chi^n_{\bf k}(q)\not=0$.
Therefore, using in addition \equ(l32b1), one obtains
$$\eqalign{
\bigl|\bigl(\Delta_p\sK_n^{-1}\bigr)(q)\PP^n_{\bf k}\bigr|
&\leq2\bigl|\Delta_p\hat \sK_n^{-1}(q)\PP^n_{\bf k}\bigr|\cr
&\leq2\bigl|t_p\hat \sK_n^{-1}(q)\PP^n_{\bf k}\bigr|\,
\bigl||\omega\cdot q|^2-|\omega\cdot(q+p)|^2\bigr|\,
\bigl|\hat \sK_n^{-1}(q)\PP^n_{\bf k}\bigr|\cr
&\leq Ck^{-\gamma}\eta^{-2n}|\omega\cdot p|,\cr
}$$
where $|\omega\cdot p|<\eta^{n+1}$ has been used again to conclude
that $|\hat \sK_n^{-1}(q)\PP^n_{\bf k}|\leq Ck^{-\gamma}\eta^{-n}$
is also verified.
This concludes the proof of bound \equ(DeltapG)
and \clm(intermediary).
\qed

\noindent
{\bf Proof of Bound \equ(bndonDpTTm).}

\noindent
Bound \equ(bndonDpTTm) is a simple consequence
of \clm(intermediary), \clm(delta1wn),
and the a priori bound
$$
||\hat P_m\Delta_pDw_m(z)||^{(m)}_{s,s'}\leq\varepsilon4^m|
\omega\cdot p|,
\equation(apDppn)
$$
valid for all $m=1,\dots,n$, $z\in B_m$ and $p$ satisfying
$|\omega\cdot p|<{1\over16}\eta^{m-1}$.
Indeed,
bounds \equ(bndinvan), \equ(boundd1Dwn) and \equ(bndtaufinal)
lead to
$$
||H_m(\tilde z)||^{(m-1)}_{s}\leq2\quad{\rm and}\quad
||\hat P_{m-1}T_m(z)||^{(m-2,m-1)}_{s,s'}\leq
\varepsilon^2r^{{m\over2}}.
\equation(jj1)
$$
Using in addition \equ(DeltapG) and the a priori bound
\equ(apDppn), one estimates that for $r=r(\eta)$ small enough,
$$
||\hat P_m\Delta_pT_m(z)||^{(m-1,m)}_{s,s'}
\leq\varepsilon^2r^{{m\over4}}|\omega\cdot p|.
\equation(jj2)
$$
Hence, \equ(bndonDpTTm) follows from
\equ(jj1), \equ(jj2) and \equ(jj3) by
taking $r=r(\eta)$ small enough and noting
that if $p$ satisfies
$|\omega\cdot p|<\eta^{m-1}/16$, then the following
estimate holds for any
operator $B\in\LL(h^{m}_s,h_{s'})$,
$$
||\hat P_mt_p B||^{(m)}_{s,s'}\leq||\hat P_{m-1}B||^{(m-1)}_{s,s'}.
\equation(veryspecial)
$$

It thus remains to check the a priori estimate \equ(apDppn).
In the sequel, we use the shorter notation $\pi_n=Dw_n$.
Using
$\Delta_p(ab)=\Delta_p ab+t_pa\Delta_pb$, one computes from the
recursive relation \equ(recrelDw) that for all $m=1,\dots,n$,
$$
\Delta_p\pi_m(z)=t_p\tilde H_m(z)\Delta_p\tilde\pi_{m-1}(\tilde z)
H_m(\tilde z)+
t_p\pi_m(z)\Delta_p\Gamma_m\pi_m(z),
\equation(recDppi)
$$
where $H_m(\tilde z)$ is given by \equ(defH), and
$$
\tilde H_m(z)=1+\pi_m(z)\Gamma_m.
\equation(deftH)
$$
To treat the first term on the right hand side of \equ(recDppi),
one first estimates, as previously, that for
$\varepsilon$ small enough,
$||H_m(\tilde z)||^{(m-1)}_s\leq2$
and, using \equ(bndinvan)
and the a priori bound \equ(aprioriDw),
$$
||\hat P_{m-1}\tilde H_m(z)||^{(m-1)}_{s'}\leq2.
\equation(bndtildeH)
$$
Next, remarking that
$\hat P_mt_p\tilde H_m=\hat P_mt_p\tilde H_mt_p\hat P_{m-1}$,
one computes
$$\eqalignno{
\hat P_mt_p\tilde H_m\Delta_p\tilde\pi_{m-1}(\tilde z)&=
\hat P_mt_p\Bigl(\tilde H_m\hat P_{m-1}t_{-p}\Delta_p\tilde\pi_{m-1}
(\tilde z)\Bigr),&\cr
&=-\hat P_mt_p\Bigl(\tilde
H_m\hat P_{m-1}\Delta_{-p}\tilde\pi_{m-1}
(\tilde z)\Bigr),
&\equat(junk1)\cr
}$$
which, with \equ(veryspecial) and \equ(bndtildeH), leads to
$$
||\hat P_mt_p\tilde H_m(z)\Delta_p\tilde\pi_{m-1}(\tilde z)
H_m(\tilde z)||^{(m)}_{s,s'}\leq
4||\hat P_{m-1}\Delta_{-p}\tilde \pi_{m-1}(\tilde z)||^{(m-1)}_{s,s'}.
\equation(firstt)
$$
In order to treat the second term on the right hand side of
\equ(recDppi), we first note that
$\Delta_p\Gamma_m=\hat P_{m-2}\Delta_p\Gamma_m\hat P_{m-2}$.
Hence, using \equ(DeltapG) and \equ(veryspecial), one estimates
that for $\varepsilon=\varepsilon(\eta)$ small enough,
$$
||\hat P_mt_p\pi_m(z)\Delta_p\Gamma_m\pi_m(z)||^{(m)}_{s,s'}
\leq\varepsilon|\omega\cdot p|.
\equation(sndt)
$$
Finally, collecting \equ(firstt) and \equ(sndt), one obtains,
with the relation
$\Delta_p\tilde\pi_{m-1}=\Delta_p\pi_{m-1}$,
$$
||\hat P_m\Delta_p\pi_m(z)||^{(m)}_{s,s'}
\leq4||\hat P_{m-1}\Delta_{-p}\pi_{m-1}(\tilde z)||^{(m-1)}_{s,s'}
+\varepsilon|\omega\cdot p|.
$$
Since $\Delta_p\pi_0=0$ for all $p\in\integer^d$,
applying the previous inequality
recursively leads to
$$
||\hat P_m\Delta_p\pi_m(z)||^{(m)}_{s,s'}
\leq\varepsilon\sum_{k=0}^{m-1}4^k|\omega\cdot p|,
$$
which finally yields \equ(apDppn).
\qed

\sectionnonr References

\ref
  \no B1
  \by Bourgain, J.
  \paper Construction of quasi-periodic solutions for Hamiltonian
  perturbations of linear equations and applications to nonlinear PDE
  \jour Internat. Math. Res. Notices, no. 11
  \vol
  \pages 475--497
  \yr 1994
\endref
\ref
  \no B2
  \by Bourgain, J.
  \paper Quasi-periodic solutions of Hamiltonian perturbations of 2D
  linear Schr\"odinger equations
  \jour Annals of Math.
  \vol 148
  \pages 363--439
  \yr 1998
\endref
\ref
  \no BGK
  \by Bricmont, J., K. Gaw\c edzki and A. Kupiainen
  \paper KAM theorem and quantum field theory
  \jour Comm. Math. Phys.
  \vol 201
  \pages 699--727
  \yr 1999
\endref
\ref
  \no CY
  \by Chierchia, L., and J. You
  \paper KAM tori for 1D nonlinear wave equations with periodic
  Boundary Conditions
  \jour Comm. Math. Phys.
  \vol 211
  \pages 497--525
  \yr 2000
\endref
\ref
  \no CW
  \by Craig, W. and E.~Wayne
  \paper Newton's method and periodic solutions of nonlinear
wave equations.
  \jour Comm. Pure Appl. Math.
  \vol 46
  \pages no. 11, 1409--1501
  \yr 1993
\endref
\ref
  \no E
  \by Eliasson, L.~H.
  \paper Perturbations of stable invariant tori for Hamiltonian
systems
  \jour Ann. Scuola Norm. Sup. Pisa Cl. Sci. (4)
  \vol 15
  \pages No. 1, 115--147
  \yr 1988
\endref
\ref
  \no G
  \by Gallavotti, G.:
  Invariant tori: a field theoretic point of view on
  Eliasson's work. In
  \book Advances in Dynamical Systems and Quantum Physics
  \publisher ed. R.~Figari, Singapore: World Scientific,
  \pages 117--132
  \yr 1995
\endref
\ref
  \no GGM
  \by Gallavotti, G., G.~Gentile and V.~Mastropietro
  \paper Field theory and KAM tori
  \jour Math. Phys. EJ
  \vol 1
  \pages No. 5, 13pp
  \yr 1995
\endref
\ref
  \no K
  \by Kuksin, S.~B.
  \paper Hamiltonian perturbations of infinite-dimensional linear
  systems with an imaginary spectrum
  \jour Funkts. Anal. Prilozh.
  \vol 21
  \pages 22--37 (1987). English transl. in Funct. Anal. Appl.
{\bf 21}, 192--205 (1988).
\endref
\ref
  \no KP
  \by Kuksin, S. and J. P\"oschel
  \paper Invariant Cantor manifolds of quasi-periodic oscillations for
  a nonlinear Schr\"odinger equation
  \jour Annals of Math.
  \vol 143
  \pages 149--179
  \yr 1996
\endref
\ref
  \no P1
  \by P\"oschel, J.
  \paper On elliptic lower dimensional tori in Hamiltonian systems
  \jour Math. Z.
  \vol 202
  \pages 559--608
  \yr 1989
\endref
\ref
  \no P2
  \by P\"oschel, J.
  \paper Quasi-periodic solutions for a nonlinear wave equation
  \jour Comment. Math. Helvetici
  \vol 71
  \pages 269--296
  \yr 1996
\endref
\ref
 \no PT
 \by P\"oschel, J. and E. Trubowitz
 \book Inverse Spectral Theory
 \publisher Academic Press, Inc.
 \yr 1897
\endref
\ref
  \no W
  \by Wayne, G.
  \paper Periodic and quasi-periodic solutions of nonlinear wave
  equations via KAM theory
  \jour Comm. Math. Phys.
  \vol 127
  \pages no. 3, 479--528
  \yr 1990
\endref
\ref
  \no Y
  \by You, J.
  \paper Perturbations of lower dimensional tori for
Hamiltonian systems
  \jour J.~Differential~Equations
  \vol 152
  \pages 1--29
  \yr 1999
\endref

\bye